\documentclass[a4paper,11pt]{article}
\usepackage[margin=2.5cm]{geometry}
\usepackage{setspace,graphicx,url,hyperref,enumitem}
\usepackage[linesnumbered,ruled]{algorithm2e}
\usepackage{amsmath,amssymb,amsfonts,amsthm,cite}
\newcommand{\N}{\ensuremath{\mathbb{N}}}
\newcommand{\R}{\ensuremath{\mathbb{R}}}

\begin{document}
\onehalfspacing
\title{Inferring the Dynamics of the State Evolution During Quantum Annealing}
\author{Elijah Pelofske, Georg Hahn, and Hristo Djidjev}
\date{Los Alamos National Laboratory}
\maketitle

\begin{abstract}
	To solve an optimization problem using a commercial quantum annealer, one has to represent the problem of interest as an Ising or a quadratic unconstrained binary optimization (QUBO) problem and submit its coefficients to the annealer, which then returns a user-specified number of low-energy solutions. It would be useful to know what happens in the quantum processor during the anneal process so that one could design better algorithms or suggest improvements to the hardware. However, existing quantum annealers are not able to directly extract such information from the processor. Hence, in this work we propose to use advanced features of D-Wave 2000Q to indirectly infer information about the dynamics of the state evolution during the anneal process. Specifically, D-Wave 2000Q allows the user to customize the anneal schedule, that is, the schedule with which the anneal fraction is changed from the start to the end of the anneal. Using this feature, we design a set of modified anneal schedules whose outputs can be used to generate information about the states of the system at user-defined time points during a standard anneal. With this process, called \textit{slicing}, we obtain approximate distributions of lowest-energy anneal solutions as the anneal time evolves. We use our technique to obtain a variety of insights into the annealer, such as the state evolution during annealing, when individual bits in an evolving solution flip during the anneal process and when they stabilize, and we introduce a technique to estimate the freeze-out point of both the system as well as of individual qubits.
\end{abstract}

\section{Introduction}
\label{sec:intro}
Quantum computers of D-Wave Systems, Inc., use a process called \textit{quantum annealing} to aim to find approximate solutions of high quality for NP-hard problems \cite{D-WaveSystems2000QuantumToday}. The type of function that the D-Wave annealer is designed to minimize is given by
\begin{align}
    Q({q_1,\ldots,q_n} )=\sum_{i=1}^n a_i q_i + \sum_{i\leq j} a_{ij} q_i q_j,
    \label{eq:hamiltonian}
\end{align}
where $n \in \N$ is the number of variables, $a_i \in \R$ are the linear weights, and $a_{ij} \in \R$ are the quadratic weights. If $q_i \in \{-1,+1\}$, eq.~\eqref{eq:hamiltonian} is called an \textit{Ising} model. If $q_i \in \{0,1\}$, it is called a quadratic unconstrained binary optimization (\textit{QUBO}) model. The two formulations are equivalent since they can be converted into each other by a linear transformation of the variables. We consider both QUBO and Ising models in this work. Many important NP-complete problems, such as the Maximum Clique, Minimum Vertex Cover, or Graph Coloring problems, can be expressed as a minimization of the type of eq.~\eqref{eq:hamiltonian} \cite{PelofskeQTOP19, Chapuis2019, PelofskeCompFront19, DwaveMapColoring, Lucas2014}.

We typically apply the following three steps in order to implement and solve an NP-hard problem on the D-Wave annealer. First, we express the problem under investigation as a minimization of the type of eq.~\eqref{eq:hamiltonian}. Second, the coefficients $a_i$ and $a_{ij}$ of the instance of eq.~\eqref{eq:hamiltonian} we want to solve are mapped onto the qubits and the connections between them (called \textit{couplers}) of the D-Wave chip \cite{TechnicalDescriptionDwave}. Third, a pre-specified number of reads (solutions) are requested from the D-Wave annealer. In particular, for each read, the value of the $i$-th variable is given as the $i$-th bit of a bitstring which D-Wave returns as its  output from the read.

\begin{figure*}
    \centering
    \includegraphics[width=0.36\textwidth]{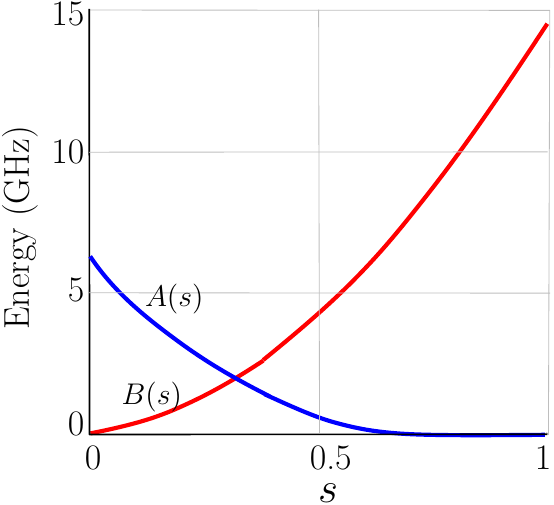}\hfill
    \includegraphics[width=0.36\textwidth]{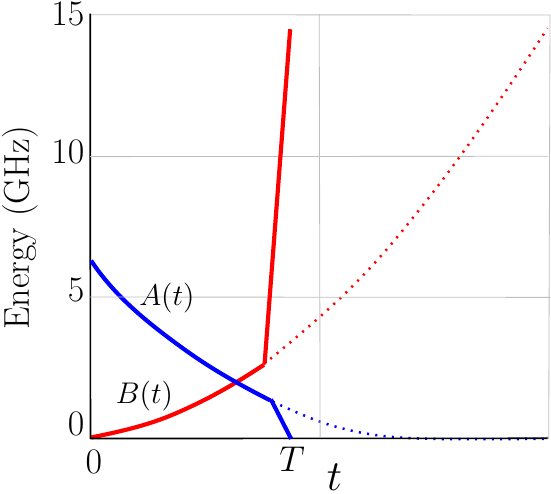}
    \caption{Left: Functions $A(s)$ (blue) and $B(s)$ (red) for D-Wave 2000Q, where $s \in [0,1]$ is the annealing fraction in a regular schedule  \cite{LANLDwave}. Right: Functions $A(s)$ and $B(s)$ in a quenching schedule.}
    \label{fig:anneal_schedule}
\end{figure*}

An operator called \textit{Hamiltonian} describes the time evolution of any quantum system. For the D-Wave processor, the following time-dependent Hamiltonian specifies its quantum system evolution:
$$H(s)=-\frac{A(s)}{2}\sum_{i=1}^n \sigma^x_i +\frac{B(s)}{2} \left( \sum_{i=1}^na_i\sigma^z_i + \sum_{i\leq j} a_{ij} \sigma^z_i \sigma^z_j \right).$$
In $H(s)$, the first term imposes an equal superposition of all states, i.e., such that each output bit string is equally likely. The actual input problem, determined by the coefficients $a_i$ and $a_{ij}$ in eq.~\eqref{eq:hamiltonian}, is encoded in the second term. A so-called \textit{anneal path} handles how the quantum system transitions from the initial quantum state to the final one, by specifying the functions $A(s)$ and $B(s)$. Figure~\ref{fig:anneal_schedule} (left) displays the values of these functions for the D-Wave 2000Q machine at Los Alamos National  Laboratory. Both functions are indexed by a parameter $s \in [0,1]$ called the \textit{anneal fraction}. Importantly, at the end of the anneal, we have $s=1$ and $A(s)=0$, meaning that the final Hamiltonian $H(1)$ is associated with a low-energy quantum system whose qubits' values can be measured to to get a high-quality solution of eq.~\eqref{eq:hamiltonian}.

The D-Wave 2000Q annealer gives more freedom to the user to tune some of the annealing control parameters \cite{dwave-technology}. Such options include a server-side spin reversal \cite{spinreversal}, pausing \cite{PowerOfPausing}, reverse annealing schedules \cite{Ohkuwa2018,Venturelli2019}, time-dependent gain in linear biases \cite{hgainQCE}, or anneal offsets \cite{Lanting2017}. Using up to $50$ user-specified time points, the \textit{anneal schedule} specifies how the anneal fraction  evolves from the start to the end of the anneal \cite[Figure~2.1]{TechnicalDescriptionDwave}. Previously, this feature has been used for improving the accuracy of the annealer by inserting a pause in the anneal schedule \cite{PowerOfPausing,PhysRevB.100.024302}.

The aim of this article is to investigate how the quantum state of the D-Wave annealer changes during an anneal process. This process is unobservable directly, since D-Wave only allows users to read off the final qubit states at the end of each anneal. For this reason, we design a set of modified anneal schedules whose outputs can be used to generate information about the states of the system at user-defined time points during a standard anneal. Using a custom anneal schedule and \textit{quenching}, a standard feature provided by D-Wave 2000Q \cite[Section~2.5.2]{TechnicalDescriptionDwave}, we follow the usual anneal curve up to a specific time at which we would like to obtain information about the state. We then modify the anneal schedule by inserting a jump to the full anneal (see Figure~\ref{fig:anneal_schedule}, right), thus getting a snapshot of the state at that intermediate time point (this idea is mentioned but not used in \cite{Amin2018}). Repeating this process for various intermediate time points allows us to \textit{slice} the anneal and then stitch all the information together in order to gain a better understanding of the dynamics of the anneal process.

We use our technique to obtain some insights into D-Wave's anneal process which, to the best of our knowledge, have not been reported previously in the literature. First, our approach allows us to visualize the state evolution during annealing, both in terms of its energy and in terms of its dynamics, i.e., we assess how volatile the measurement of each individual qubit of the lowest energy state is during annealing. We repeat our experiments for a varying number of slices and annealing times.

Second, in order to better understand what happens during an anneal, we present a simple genetic optimization scheme \cite{MitchellGeneticAlgorithms} designed to find a QUBO that benefits greatly from the quantum annealing cycle, in the sense that it exhibits substantial improvements in terms of energy decrease. We contrast the (optimized) QUBO obtained in this fashion with a random QUBO. Moreover, we study the characteristics and best parameter choices for our genetic algorithm. Genetic algorithms have previously been used in connection with quantum annealing, though in the context of hybrid quantum-classical solvers \cite{King2019}.

Third, our technique gives us a new way to determine an estimate of the \textit{freezeout point}, a hypothetical point that is defined as an anneal fraction $s$ significantly smaller than $1$, after which the dynamics essentially stalls, and which can be found by fitting a Boltzmann distribution to the annealer's output. While we show that a freezeout point so defined may not always exist or be identifiable by the existing methods, we demonstrate that our slicing method allows us to determine an analogue we call a \textit{quasi-freezeout point (QFP)}. It is defined as a point at which either the energy does not significantly improve anymore or, at an individual qubit level, a point at which the state/value of each individual qubit stays fixed. 
For the former, we analyze the energy values determined by the slicing method and determine the point after which the slope of the energy plot becomes close to $0$. We compare our approach to the one based on estimating the parameters of a Boltzmann distribution, e.g., \cite{Benedetti2016}.
For the latter, qubit-level QFP, we slice the anneal at various stages (for instance, using $1000$ slices for a $1000$ microsecond anneal), and observe how each qubit's measured value ($+1$ or $0/-1$) changes over the course of the anneal. The QFP of a qubit can then be defined as the approximate location at which the measured value of the qubit stays invariant until the end of the anneal. Early works on measuring the freezeout of a quantum system by means of a Boltzmann distribution \cite{Chancellor2016scirep} or the Kibble-Zurek mechanism \cite{Chancellor2016arxiv} are available in the literature.

The article is structured as follows. Section~\ref{sec:methods} presents our approach to slice the anneal process, that is, the anneal schedule we employ in order to quench the anneal process at any intermediate time. We also present our genetic algorithm to find a QUBO which exhibits a substantial state evolution during the anneal, both in terms of the total number of bit flips of all involved qubits, as well as the total energy decrease during the anneal. Using that QUBO, Section~\ref{sec:simulations} will visualize the state evolution during the anneal, both in terms of the number of bit flips and the total energy change. We investigate how a random QUBO compares to our optimized one, and how the number of slices influences the results. Finally, we look at experimental measurements of the freezeout point for various examples. A summary and discussion of our methodology and results can be found in Section~\ref{sec:discussion}.

This paper is a journal version of an article presented at the \textit{20th International Conference on Parallel and Distributed Computing, Applications and Technologies (PDCAT) 2019} \cite{pdcat}. Extending the conference version, the present paper also considers the estimation of the freezeout point based on the estimation of effective temperatures (Section~\ref{sec:measuring_freezeout}), information collected during slicing (Section~\ref{sec:spline}), and we consider schedules that use a pause during the anneal process (Section~\ref{sec:pausing}). Additionally, the present paper contains more extensive experiments investigating the effect of a pause in the anneal process (Section~\ref{sec:evolution_pausing}), and the slicing of a QUBO for the Maximum Clique problem (Section~\ref{sec:optimizedMaxClique}), a well-known NP-hard problem. Since the connectivity structure (the quadratic couplers) of the function of eq.~\eqref{eq:hamiltonian} being minimized on D-Wave does not typically match the one of the physical qubits on the D-Wave chip, one important aspect of quantum annealing pertains to the computation of a suitable minor embedding of the problem structure in eq.~\eqref{eq:hamiltonian} onto the physical qubits. In this process, logical qubits are often represented as a set of physical ones that are supposed to act as one (called a "chain"). Even though chains of physical qubits represent the same logical qubit, it is not guaranteed that all of them take the same value after readout at the end of the anneal. This is called a "broken chain". We visualize how chain breaks occur during the anneal (Section~\ref{sec:chainbreaks}), and estimate the freezeout point for all individual qubits (Section~\ref{sec:freezeout_chimera}). Finally, most of the figures are new or revised, compared to the conference version.

\section{Methodology}
\label{sec:methods}
This section, which presents the methodology underlying our work, consists of two parts. First, Section~\ref{sec:slicing} describes the quenching technique we use in order to get information about the state evolution during the anneal process. To visualize results later in the simulations, we are interested in finding a QUBO that exhibits a pronounced evolution during the anneal process, that is, whose lowest energy result from a 1-microsecond anneal is (significantly) greater than that of a longer (1000-microsecond) anneal. We find such a QUBO with the help of a genetic algorithm presented in Section~\ref{sec:genetic_algo}. This section ends with a review of a published technique to estimate freezeout points using the technique of \cite{Benedetti2016} in Section~\ref{sec:measuring_freezeout}, and a short discussion of how to slice an anneal schedule with a pause in Section~\ref{sec:pausing}.

\subsection{Slicing the anneal process}
\label{sec:slicing}
\begin{figure}[t]
    \centering
    \includegraphics[width=0.45\textwidth]{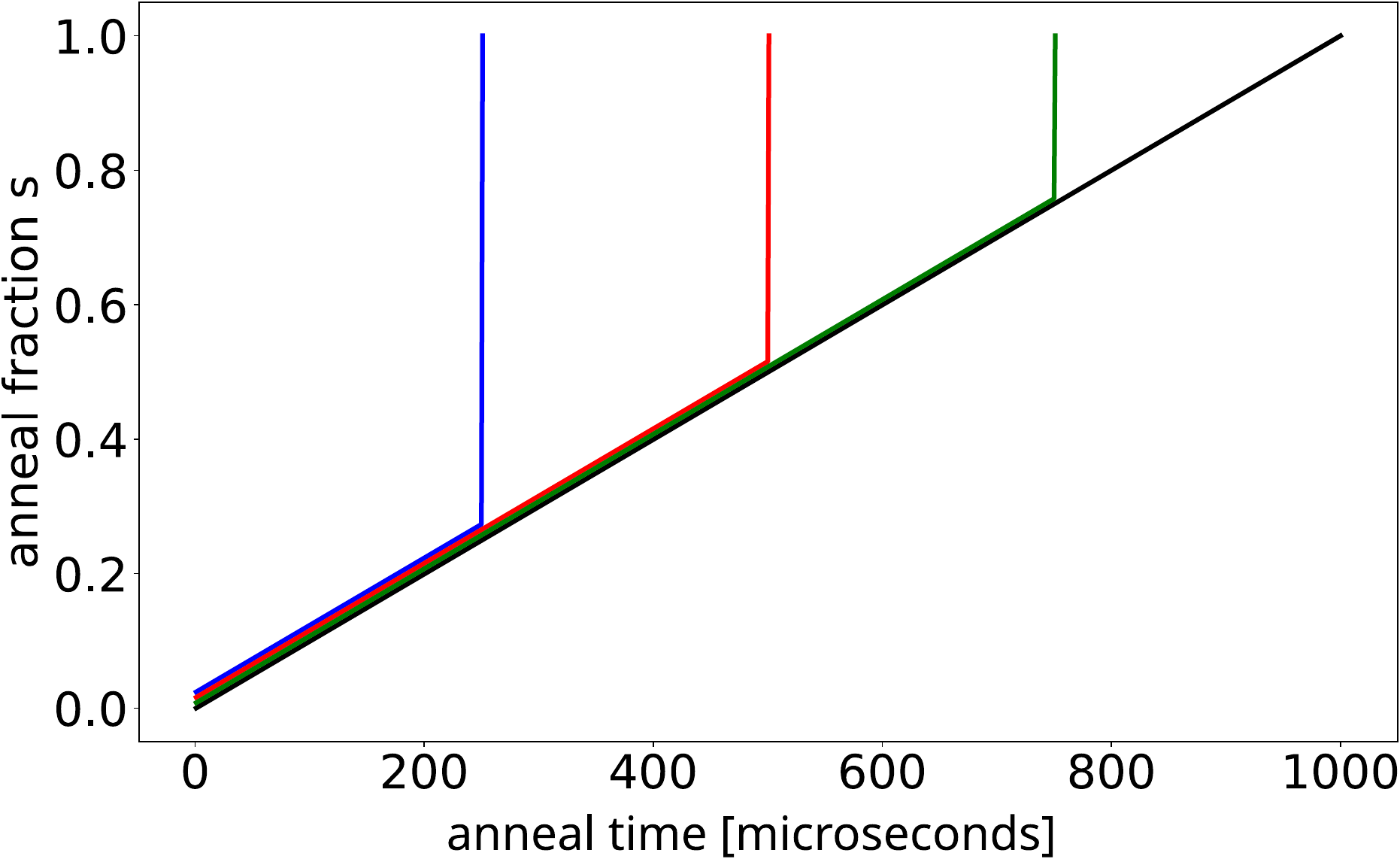}
    \caption{Anneal schedule with quenching (near-vertical jumps) at time points 250 (blue), 500 (red), and 750 (green) microseconds. Total anneal time of 1000 microseconds. The full anneal schedule is shown by the black line.}
    \label{fig:quenching}
\end{figure}

The goal of this work is to infer information on how the states evolve during the anneal process of D-Wave 2000Q. For this, we use a feature recently introduced on D-Wave which allows users to manually define an anneal schedule by specifying up to $50$ points on the anneal curve. In particular, we make use of a feature called "quenching" that allows us to jump to the full anneal fraction of $1$ within one microsecond at any point in the anneal process (for the precise conditions under which the quench is possible, see \cite{TechnicalDescriptionDwave}).

Figure~\ref{fig:quenching} illustrates the sequences of quenches we employ. For the $i$-th quench used to slice the anneal process at time $t_i$, we first follow the standard anneal curve connecting the point $P_0=(0,0)$ (i.e., $t=0$ and $s=0$) with the point $P_1=(t_i,t_i/T)$, where $T$ is the full anneal duration. Ideally, the jump from the intermediate time $t_i$ to the end of the anneal has to be done as quickly as possible, i.e., we would like to have $P_2=(t_i,1)$, in order to ``freeze" the current state at $t_i$. However, the hardware constraints mentioned above do not allow such a quench, and we have to jump to the full anneal fraction of $s=1$ at the next time point $t_i+1$, thus connecting $P_1$ with $P_2=(t_i+1,1)$.

Though the jump we employ is not perfectly vertical (our anneal curve jumps to $1$ at $t_i+1$), we expect this jump to not considerably change the result. But we also would like to reduce the bias caused by the $1 \mu s$ quench, since for some problems a $1 \mu s$ anneal could reduce noticeably the energy of the current state. Hence, in the next subsection, we discuss our approach to choosing problems for which the distortion caused by the quench is significantly reduced.

\subsection{A genetic algorithm for constructing a suitable QUBO/Ising model}
\label{sec:genetic_algo}
It is known that different QUBO/Ising models pose different levels of difficulty to the D-Wave annealer in terms of finding a high quality solution \cite{Chapuis2019}. In particular, random Ising models might pose rather simple problems to a quantum annealer \cite{Katzgraber2014}. In our experiments presented in Section~\ref{sec:simulations}, we aim to investigate the state evolution during the anneal process, and therefore seek to solve a non-trivial QUBO/Ising instance for which the effect of quenching is as small as possible. Additionally, we seek a QUBO/Ising instance which benefits greatly from the quantum annealing cycle in that it also exhibits a distinctive evolution during the anneal process, measured both in terms of energy decrease and bit flips.

\begin{algorithm}[t]
    \caption{Finding a suitable QUBO/Ising model for slicing\label{algo:genetic}}
    \SetKwInOut{Input}{input}
    \SetKwInOut{Output}{output}
    \SetKwFor{Repeat}{repeat}{}{end}
    \Input{$N$, $p_\text{cross}$, $p_\text{mut}$, $R$\;}
    \Output{optimized instance for slicing\;}
    $S \leftarrow \{Q_1,\ldots,Q_N \}$, where each $Q_i$ contains all linear  and quadratic terms of the hardware connectivity graph\;
    \label{line:init}
    \For{\textnormal{\textbf{each}} $Q \in S$}{
        Set each individual linear coefficient of $Q$ to a random value in $(-2,2)$\;
        Set each quadratic coefficient of $Q$ to a random value in $(-1,1)$\;
    }
    \Repeat{$R$ \textnormal{\textbf{times}}}{
        \For{\textnormal{\textbf{each}} $Q \in S$}{
            \label{line:fitness_loop}
            Across $1000$ anneals, find energy difference $\Delta$ between the $1\%$ minimum energy samples for $Q$ for a $1$ microsecond and $1000$ microsecond full anneal\;
            Set fitness $f_Q \leftarrow \Delta$\;
            \label{line:fitness}
        }
        $F_0 \leftarrow$ proportion $p_\text{cross}$ of largest $f_Q$ values\;
        $S_0 \leftarrow \{Q \in S: f_Q \in F_0\}$\;
        $S_1 \leftarrow \emptyset$\;
        \Repeat{$N$ \textnormal{\textbf{times}}}{
            \label{line:crossover}
            Draw two random $Q_1,Q_2 \in S_0$ and select coefficients randomly from either $Q_1$ or $Q_2$ with probability $0.5$; store new problem instance in $S_1$\;
        }
        $S_2 \leftarrow \emptyset$\;
        \For{$Q \in S_1$}{
            \label{line:mutation}
            Set $Q' \leftarrow Q$ and overwrite each coefficient in $Q'$ independently with probability $p_\text{mut}$ (for linear coefficients use random value in $(-2,2)$, for quadratic coefficients use $(-1,1)$)\;
            Add $Q'$ to $S_2$\;
        }
        $S \leftarrow S_2$\;
    }
    \Return{\textnormal{fittest $Q \in S$ as defined in line~\ref{line:fitness}}}\;
\end{algorithm}

To find such a problem instance, we employ the genetic algorithm presented in Algorithm~\ref{algo:genetic}, which works as follows. The algorithm starts by initializing a population of $N \in \N$ problem instances (QUBO or Ising models), which are collected in a set $S$. Those instances are randomly generated as follows: each problem instance maps onto the entire Chimera graph, the graph encoding the connectivity structure of the physical qubits on the D-Wave chip, see \cite{TechnicalDescriptionDwave}. Its linear coefficients $a_i$ are independently sampled from $(-2,2)$, and its quadratic coefficients $a_{ij}$ are independently sampled from $(-1,1)$.

Then, the algorithm proceeds by evaluating the fitness of the current population of problem instances. To this end, for each $Q \in S$, we find the mean of the best 1 percent of energies in a $1$ microsecond and a $1000$ microsecond anneal. We set the fitness $f_Q$ for each $Q \in S$ to $|f_{Q,1}-f_{Q,1000}|$, where $f_{Q,1}$ is the average of the best $1\%$ samples from the 1 microsecond anneal, and $f_{Q,1000}$ is the average of the  best $1\%$ samples from the 1000 microsecond anneal. Using the mean of the best $1\%$ samples results in more stable results and reduces the effect of the noise, compared to merely considering the minimum energy. The fitness $f_Q$ is, at the same time, the objective function that the genetic algorithm optimizes. This ensures that Algorithm~\ref{sec:genetic_algo} will optimize for problem instances having the property that between a $1$ microsecond and a full $1000$ microsecond anneal, the state evolves considerably in terms of the energy.

Next, the $p_\text{cross}$ portion of fittest individuals are selected from the population for cross-over and mutation. Those problem instances are stored in the set $S_0$. Then we restore the original size $N$ of the population by crossing the fittest individuals from set $S_0$: we randomly choose two instances $Q_1$ and $Q_2$ from $S_0$ and generate a new one by selecting each individual linear and quadratic coefficient independently from either $Q_1$ or $Q_2$ with probability $0.5$. We store the new instance in $S_1$ and repeat this step $N$ times.

Finally, a mutation step is applied to the new population. Each instance $Q \in S_1$ from the new population is first copied into $Q' \leftarrow Q$. Then, we overwrite any of the coefficients of $Q'$ with a probability $p_\text{mut}$. For those coefficients which are being overwritten, a linear coefficient is sampled randomly from $(-2,2)$, and a quadratic coefficient is sampled from $(-1,1)$. The resulting instances are stored in a new set $S_2$. After setting $S \leftarrow S_2$, the genetic algorithm is restarted with the newest population.

The entire process is repeated over $R$ iterations. After the last iteration, we return the fittest $Q \in S$ as the result of our algorithm, where $S$ is the newest population generation and fitness is calculated as $f_Q$.

The dependence of the genetic algorithm described in Section~\ref{sec:genetic_algo} on its parameters is evaluated in Section~\ref{sec:GA_parameter_choice}, where we also suggest default parameter choices.

\subsection{Estimating the freezeout point}
\label{sec:measuring_freezeout}
The slicing approach we introduce in Section~\ref{sec:slicing} allows one to track how the current state, in particular its associated energy, evolves during the anneal process. A natural parameter to look at in this scenario is the so-called \textit{freezeout point} \cite{PhysRevA.92.052323, PhysRevApplied.8.064025}, defined as a point before the end of the anneal at which the evolution of the quantum state ``freezes'', and after which not much quantum dynamics occurs except for fluctuations due to noise. While such a freezeout point may not always be well-defined theoretically or easy to compute, our slicing method gives an alternative way to the literature for looking at the evolution of the anneal process. Overall, we are interested in determining a point where the dynamics of the anneal process slows down or stops \cite{Raymond2016}.

In order to be able to make a comparison, we want to compute a freezeout point estimate based on an established technique such as \cite{PhysRevA.92.052323} or \cite{PhysRevApplied.8.064025}, and then compare it with the freezeout point estimate computed by our method.

We use the statistical methodology of \cite{Benedetti2016} to estimate the freezeout point. This method is based on the Boltzmann distribution, the theoretical distribution of ground-state energies of the annealer. Briefly, estimating a certain parameter of the Boltzmann distribution, the effective inverse temperature $\beta_\text{eff}$ of the annealer, allows one to estimate the freezeout point. Denoting $\beta=\beta_\text{eff}$, we first observe that the (thermodynamic) probability of observing an energy $E$ is given by $P_\beta(E) = g(E) \exp(-\beta E)/Z(\beta)$, where $g(E)$ is the degeneracy of the energy level $E$ and $Z(\beta)$ is the partition function serving as normalization factor.

Second, for two energies $E_1$ and $E_2$, the log ratio of probabilities is
$$l(\beta) := \log \frac{P_\beta(E_1)}{P_\beta(E_2)} = \log \frac{g(E_1)}{g(E_2)} - \beta \Delta E.$$
The main idea of \cite{Benedetti2016} is to rescale all coefficients of the Hamiltonian under consideration by a factor $x \in (0,1)$. Letting $\beta'=x \beta_\text{eff}$, computing the difference of log probabilities allows us to eliminate the unknown degeneracies $g(\cdot)$ since
\begin{align}
    \Delta l = \frac{P_\beta(E_1) P_{\beta'}(E_2)}{P_\beta(E_2) P_{\beta'}(E_1)} = \Delta \beta \Delta E.
    \label{eq:perdomo}
\end{align}
The expression in eq.~\eqref{eq:perdomo} allows us to compute an empirical freezeout point estimate. Indeed, for a fixed $x \in (0,1)$, we rescale our Hamiltonian under consideration and obtain a fixed number of $R$ anneals for both the unscaled (original) and rescaled Hamiltonians. By binning the obtained energies of those $R$ anneals per Hamiltonian, we can compute empirical probabilities for each energy bin. Using $K=\lceil \sqrt{2R} \rceil$ bins is recommended in \cite{Benedetti2016}. Afterwards, we can select all pairs of bins for both Hamiltonians and compute both $\Delta l$ and $\Delta E$ per pair. According to eq.~\eqref{eq:perdomo}, the data pairs $(\Delta l,\Delta E)$ will lay on a straight line through the origin with slope $\Delta \beta = (x-1)\beta_\text{eff}$. Estimating that slope thus immediately allows us to compute $\beta_\text{eff}$.

Having computed $\beta_\text{eff}$, the freezeout point estimate follows from a simple calculation. We observe that using the operating temperature of D-Wave 2000Q of $T=15$ mK, and the Boltzmann constant $k_B = 20.83661$ GHz/K, we can convert $\beta_\text{eff}$ via $\beta_\text{eff} k_B T$ into a point on the anneal schedule $B(s)$ displayed in Fig.~\ref{fig:anneal_schedule}. Finding $s^\ast$ such that $B(s^\ast)=\beta_\text{eff} k_B T$ yields an estimate of the freezeout point.

The aforementioned method to measure the freezeout point is dependent on the choice of the parameter $x \in (0,1)$ used to rescale the Hamiltonian. Typically, values of $x$ too close to $0$ will cause the rescaled Hamiltonian to be too different from the original one (thus also causing its Boltzmann distribution to be too far away from the one of the unscaled Hamiltonian), whereas values of $x$ too close to $1$ will not yield a good separation of the rescaled and unscaled energy measurements used in eq.~\eqref{eq:perdomo}. We typically choose $x$ on a grid in the interval $[0.6,0.95]$. After computing the freezeout point estimate for each such $x$, in the experiments of Section~\ref{sec:simulations} we report the measurement of the freezeout point corresponding to the largest value of $x<1$.

\subsection{Slicing an anneal schedule with a pause}
\label{sec:pausing}
\begin{figure}
    \centering
    \includegraphics[width=0.45\textwidth]{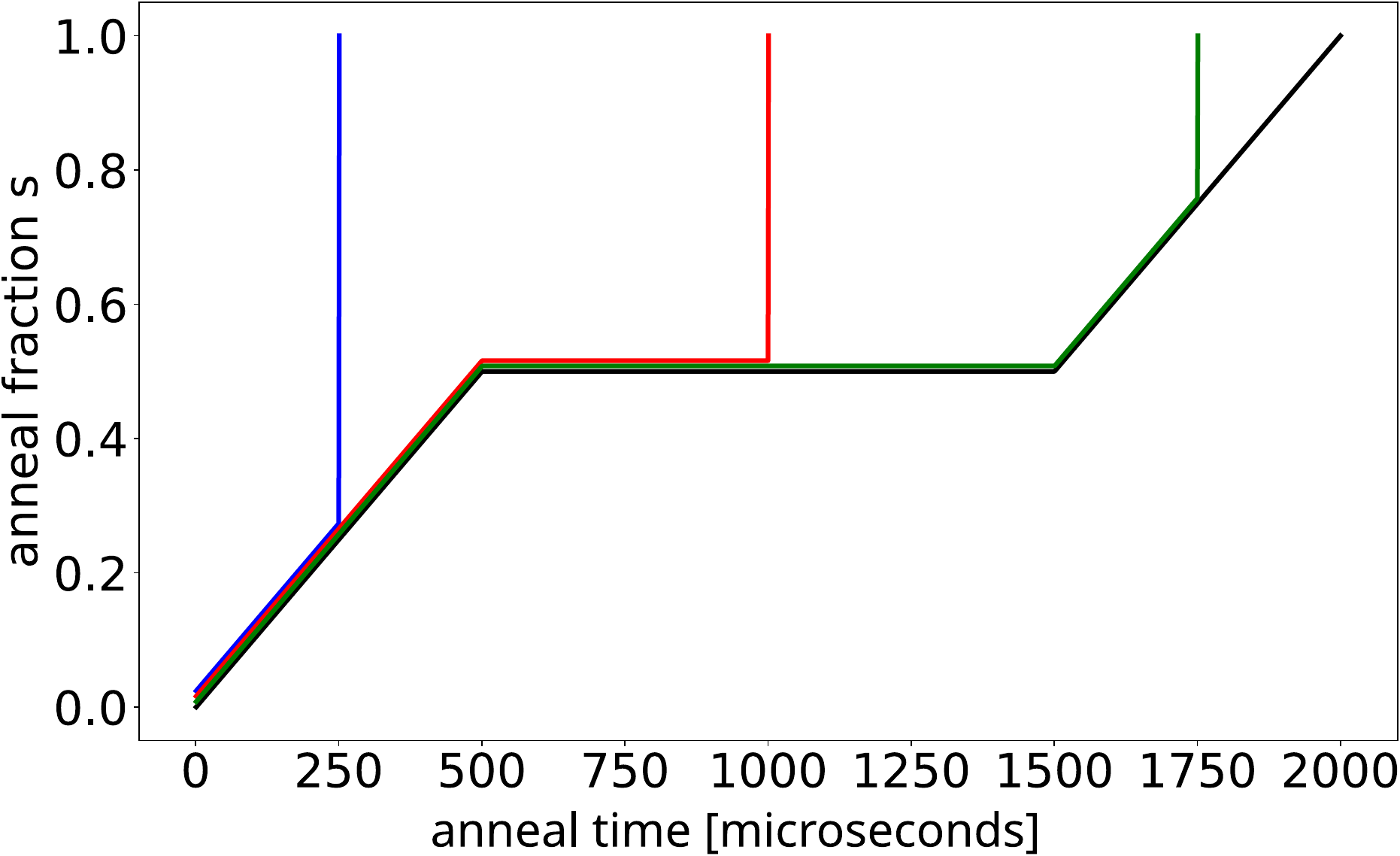}
    \caption{Anneal schedule with a pause of length 1000 microseconds inserted 500 microseconds into the anneal. Quenching (near-vertical jumps) at time points 250 (blue), 1000 (red), and 1750 (green) microseconds. The full anneal schedule is shown by the black line. Total anneal time of 2000 microseconds.}
    \label{fig:pausing_schedule}
\end{figure}

In the experiments of Section~\ref{sec:simulations}, we also aim to investigate the effect of inserting a pause into the anneal process. Fig.~\ref{fig:pausing_schedule} illustrates an annealing schedule that includes a pause (the black line), and a series of slicing schedules (green, red, and blue lines).
Specifically, the annealing with pause schedule in Fig.~\ref{fig:pausing_schedule} starts with 500 microseconds regular anneal, at which a pause of 1000 microseconds is inserted. Afterwards, the normal anneal schedule is resumed until it reaches the total anneal time of 2000 microseconds. Slicing the anneal before and during the pause allows us to observe the state evolution even while the Hamiltonian evolution is paused.

\section{Experiments}
\label{sec:simulations}
This section presents our experimental setup and results. We start with an assessment of the tuning parameters of our genetic algorithm (Section~\ref{sec:GA_parameter_choice}). In Section~\ref{sec:random_vs_optimized}, we present first results on the slicing technique of Section~\ref{sec:slicing}, demonstrating that the Ising model returned by Algorithm~\ref{algo:genetic} indeed yields more pronounced changes during the anneal. We then focus on the evolution of the energy (Section~\ref{sec:evolution_energy}) and the Hamming distance (Section~\ref{sec:evolution_hamming}) during annealing.

Moreover, we look into a particular feature of the D-Wave 2000Q, the inclusion of a pause in the anneal process (Section~\ref{sec:evolution_pausing}). Since we are also interested in slicing chained problems of practical importance, we modify Algorithm~\ref{algo:genetic} to yield a QUBO for a well-known NP-hard problem, the Maximum Clique problem. We use this QUBO to investigate how the energy, Hamming distance, and freezeout point estimate behave (Section~\ref{sec:optimizedMaxClique}). Our slicing technique allows us to define a quantity related to the freezeout point which we define in Section~\ref{sec:spline}. Additionally, it allows us to visualize how chain breaks occur during the anneal (Section~\ref{sec:chainbreaks}), and we investigate individual freezeout estimates for all qubits (Section~\ref{sec:freezeout_chimera}).

Apart from Section~\ref{sec:random_vs_optimized}, which also displays results for a random Ising model, all figures in the remainder of the simulations were computed using an Ising model obtained with the genetic algorithm described in Section~\ref{sec:genetic_algo}, or a modification thereof.

\subsection{Parameter choices for the genetic algorithm}
\label{sec:GA_parameter_choice}

\begin{figure*}[t]
    \centering
    \includegraphics[width=0.49\textwidth]{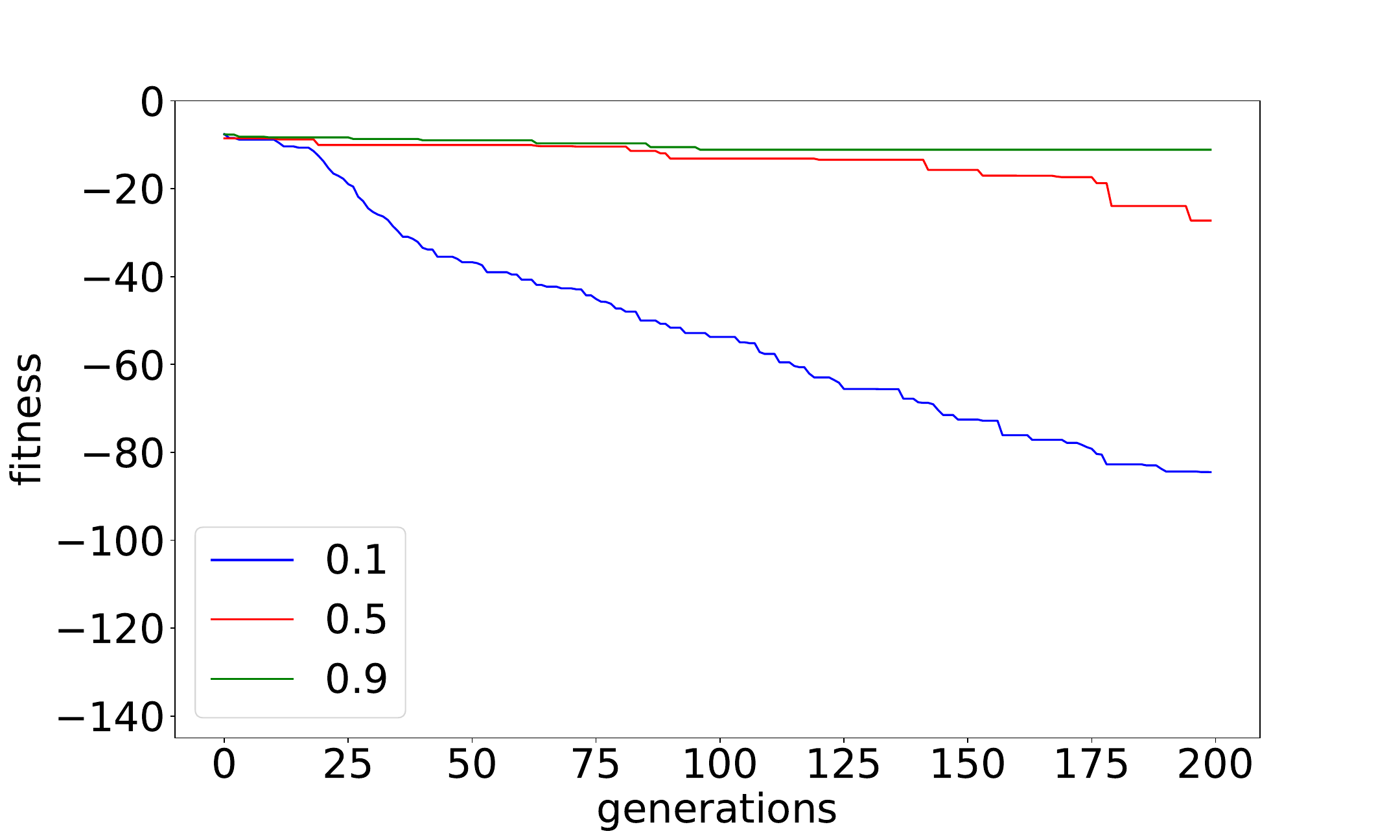}\hfill
    \includegraphics[width=0.49\textwidth]{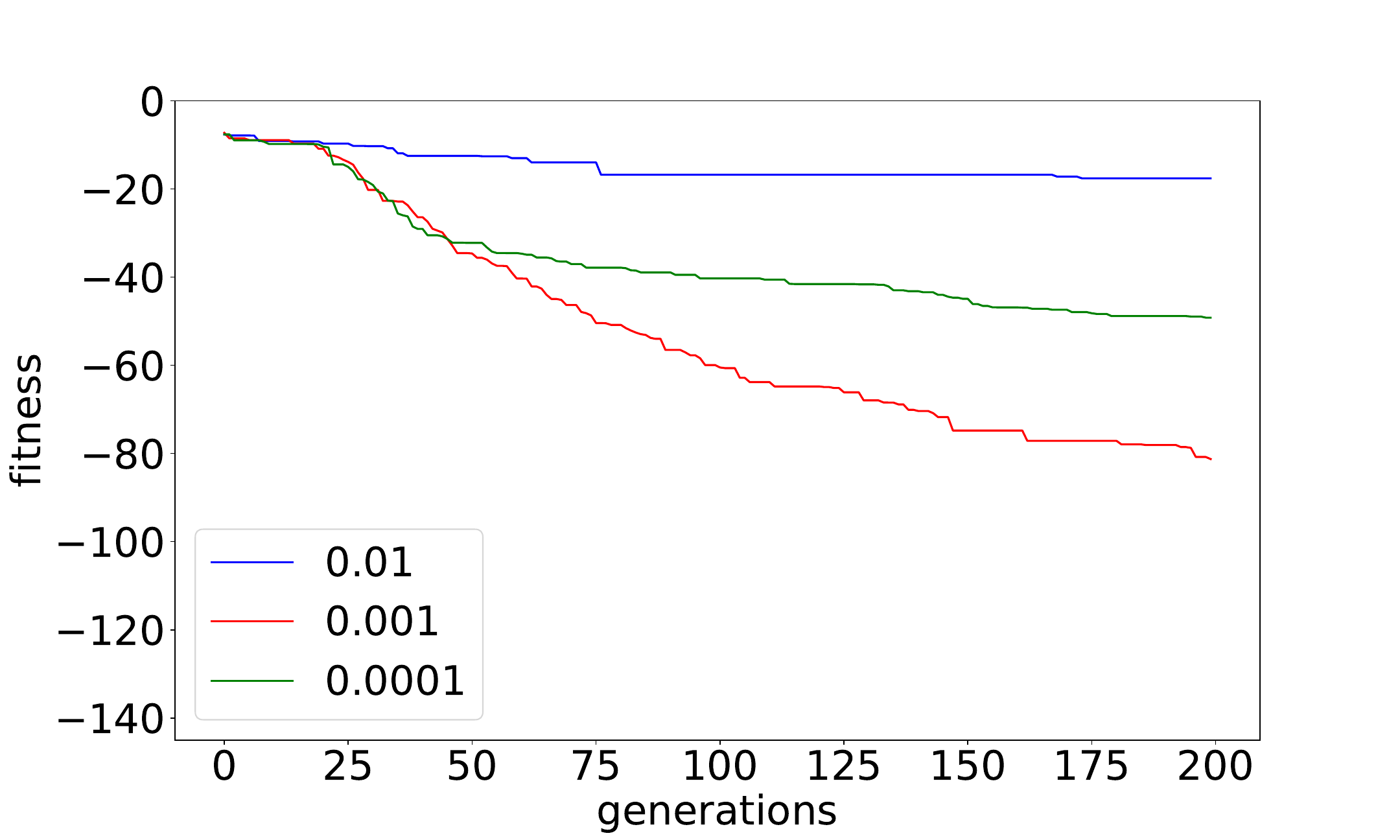}\\
    \includegraphics[width=0.49\textwidth]{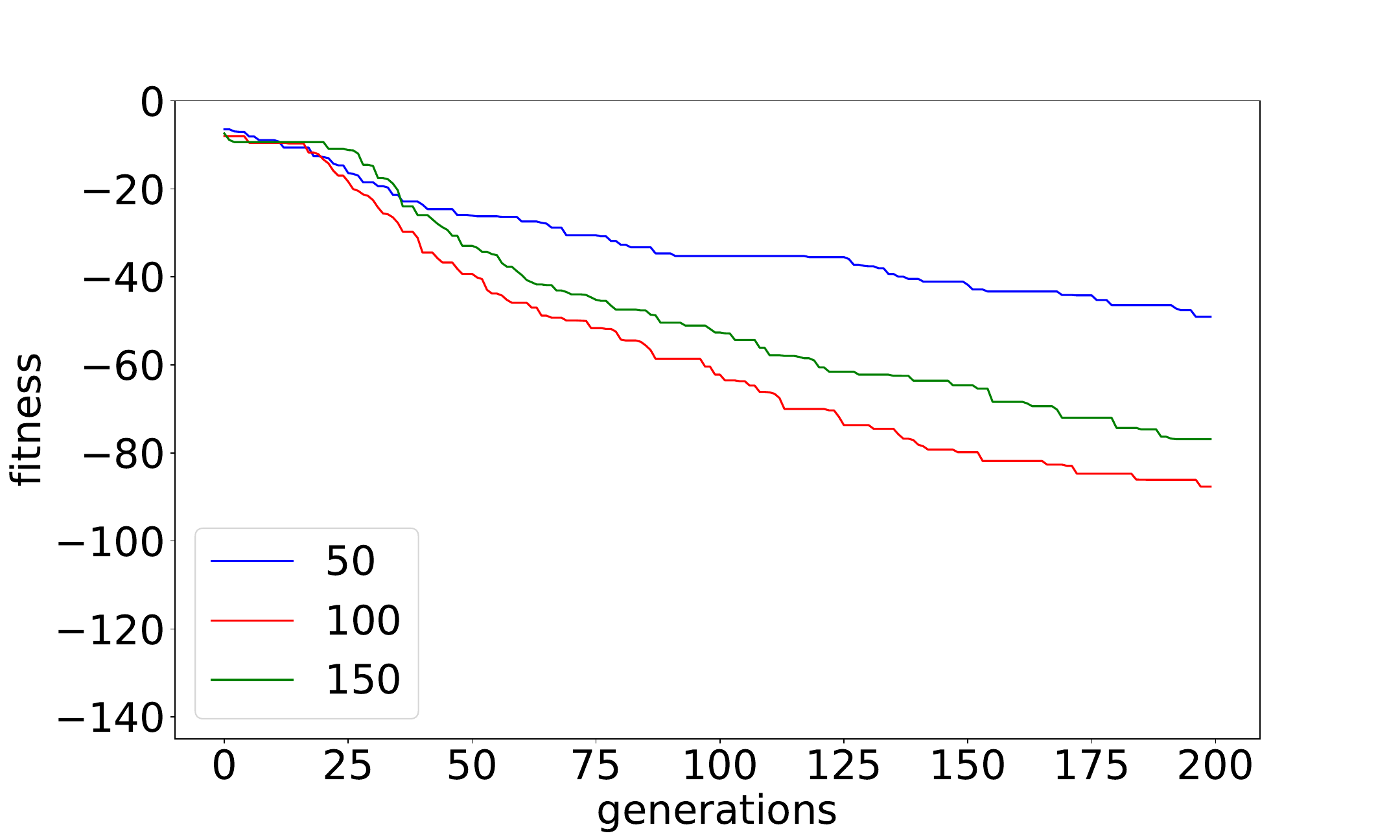}\hfill
    \includegraphics[width=0.49\textwidth]{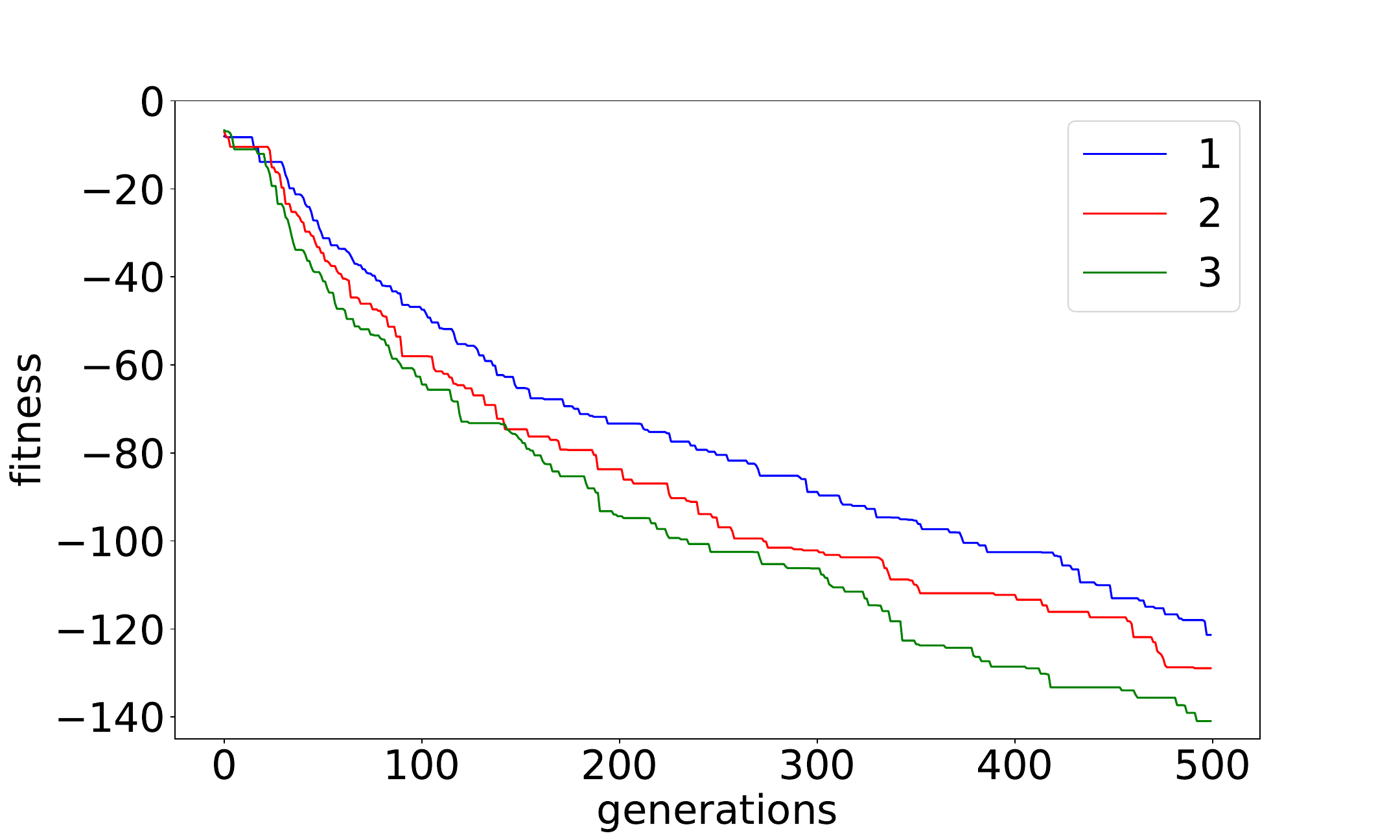}
    \caption{Dependence of the genetic algorithm on its tuning parameters, while keeping the other parameters at their default values, measured in the fitness $f_Q$ (lower values are better). Crossover proportion $p_\text{cross} \in \{0.1,0.5,0.9\}$ (top left), mutation rate $p_\text{mut} \in \{0.0001,0.001,0.01\}$ (top right), and population size $N \in \{50,100,150\}$ (bottom left). Bottom right shows three runs of the genetic algorithm with the optimized parameters.}
    \label{fig:genetic_algo}
\end{figure*}

We start by assessing the dependence of the genetic algorithm described in Section~\ref{sec:genetic_algo} on its tuning parameters. For this we first define a set of default parameters: population size $N=100$, crossover proportion $p_\text{cross}=0.1$, and mutation rate $p_\text{mut}=0.001$.

While keeping two of the default parameters fixed, we vary the third parameter in Figure~\ref{fig:genetic_algo}. We report the fitness $f_Q$ in all plots. As can be seen from the definition of $f_Q$ as the energy difference between the $1\%$ minimum energy samples for a $1$ microsecond and a $1000$ microsecond full anneal (see Section~\ref{sec:genetic_algo}), the lower the fitness value the better (i.e., more pronounced) the energy evolution during the anneal.

Figure~\ref{fig:genetic_algo} shows that selecting a low crossover rate ($p_\text{cross}=0.1$) seems to be advantageous. However, choosing the population size or the mutation rate to be too low or too high is disadvantageous, thus leading us to the choices $N=100$ and $p_\text{mut}=0.001$. The default parameters will be used in the remainder of the simulations. Figure~\ref{fig:genetic_algo} (bottom right) shows three runs of the genetic algorithm with default parameters for $500$ generations, demonstrating a stable behavior of the tuned algorithm. Whenever we apply Algorithm~\ref{algo:genetic} in the following sections, we will always employ $200$ generations (iterations).

\subsection{Evolution of the quantum state for two types of Ising models}
\label{sec:random_vs_optimized}
We verify that using an Ising model computed with the genetic algorithm is indeed advantageous for investigating the evolution of the quantum state during the anneal. We refer to the optimized Ising model returned by Algorithm~\ref{algo:genetic} as simply the \textit{Chimera Ising} in the remainder of the text. To be precise, we employ the Ising model with best fitness from the third run of the genetic algorithm which is displayed in the lower right panel in Figure~\ref{fig:genetic_algo} as a green line.

\begin{figure}
    \centering
    \includegraphics[width=0.49\textwidth]{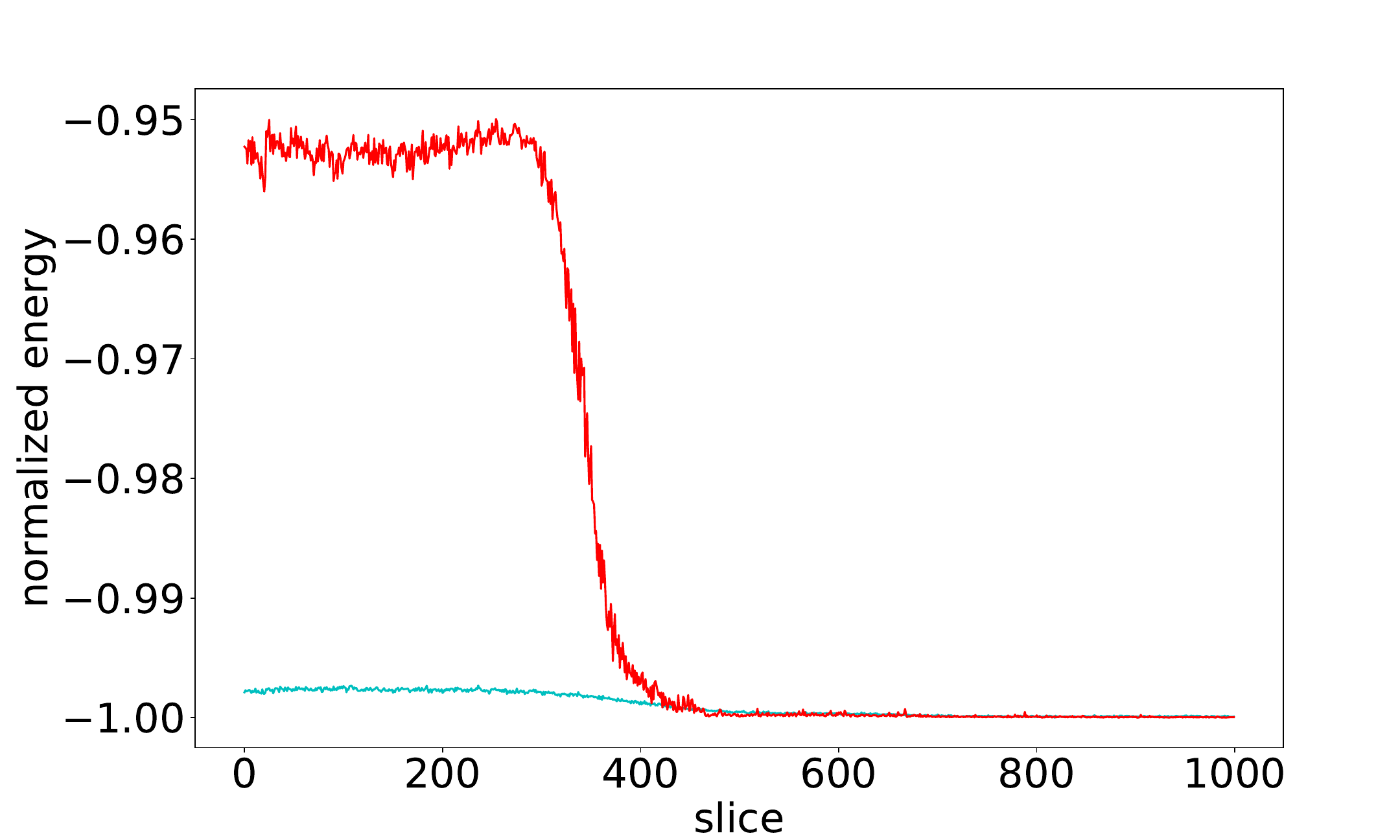}
    \caption{Averages of the minimum 1 percent energies found in 10 runs of 1000 anneals. Randomly generated and unstructured Ising model (cyan) and the Chimera Ising model (red). Both datasets are normalized by dividing each value by their respective minimum value.}
    \label{fig:1000_rand_opt}
\end{figure}

Figure~\ref{fig:1000_rand_opt} shows, for $1000$ slices (i.e., a $1000$ microsecond anneal), the progression of the average of the $1$ percent minimum energies found. We observe that the total change in energy is considerably more pronounced for the Chimera Ising returned by Algorithm~\ref{algo:genetic} compared to a randomly generated non-optimized and unstructured Ising. Looking at the plot it seems straightforward to define the freezeout point to occur at roughly slice $500$ where both curves stabilize horizontally. We will investigate the relationship between the point at which the energies stabilize in the slicing plots and the freezeout point estimate computed with the methodology of \cite{Benedetti2016} (see Section~\ref{sec:measuring_freezeout}) in more detail in the next subsection.

\subsection{Energy evolution during the anneal process}
\label{sec:evolution_energy}
We now look at the evolution of energies for the Chimera Ising model. In this and all following sections, we always slice an anneal process using 1000 and 2000 slices. Since we slice in steps of one microsecond, the figures showing 1000 slices are always computed for a total anneal time of 1000 microseconds, and the figures showing 2000 slices have a total anneal time of 2000 microseconds. For each slice we run $1000$ anneals on D-Wave 2000Q. In the plots, we report both the mean energy among all samples returned by D-Wave 2000Q as well as the average best $1\%$ energies, that is the mean among the $1\%$ lowest energies observed among the $1000$ samples. The Hamming distance we report is the average distance among the bitstrings corresponding to the $1\%$ best (lowest energy) samples.

\begin{figure*}
   \centering
    \includegraphics[width=0.49\textwidth]{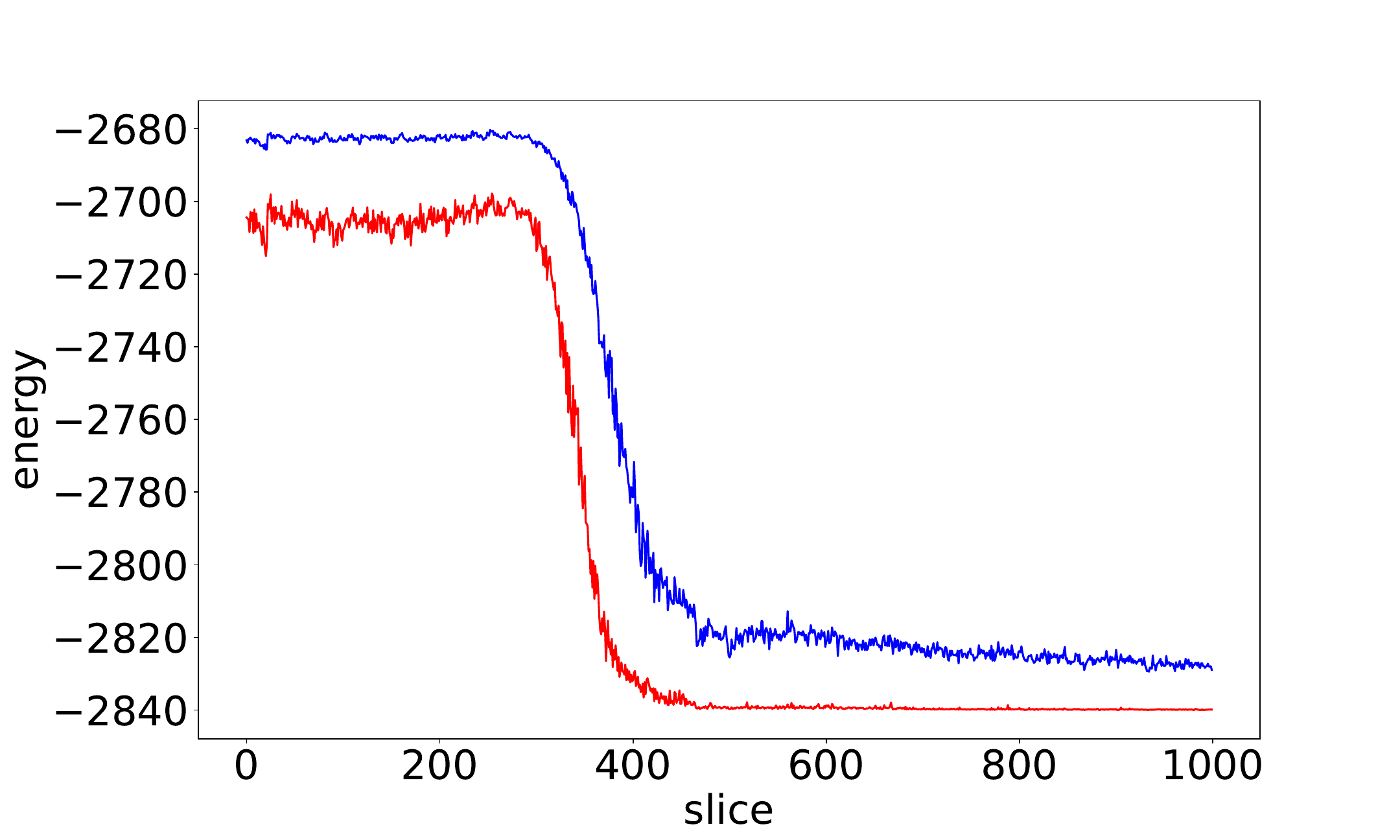}\hfill
    \includegraphics[width=0.49\textwidth]{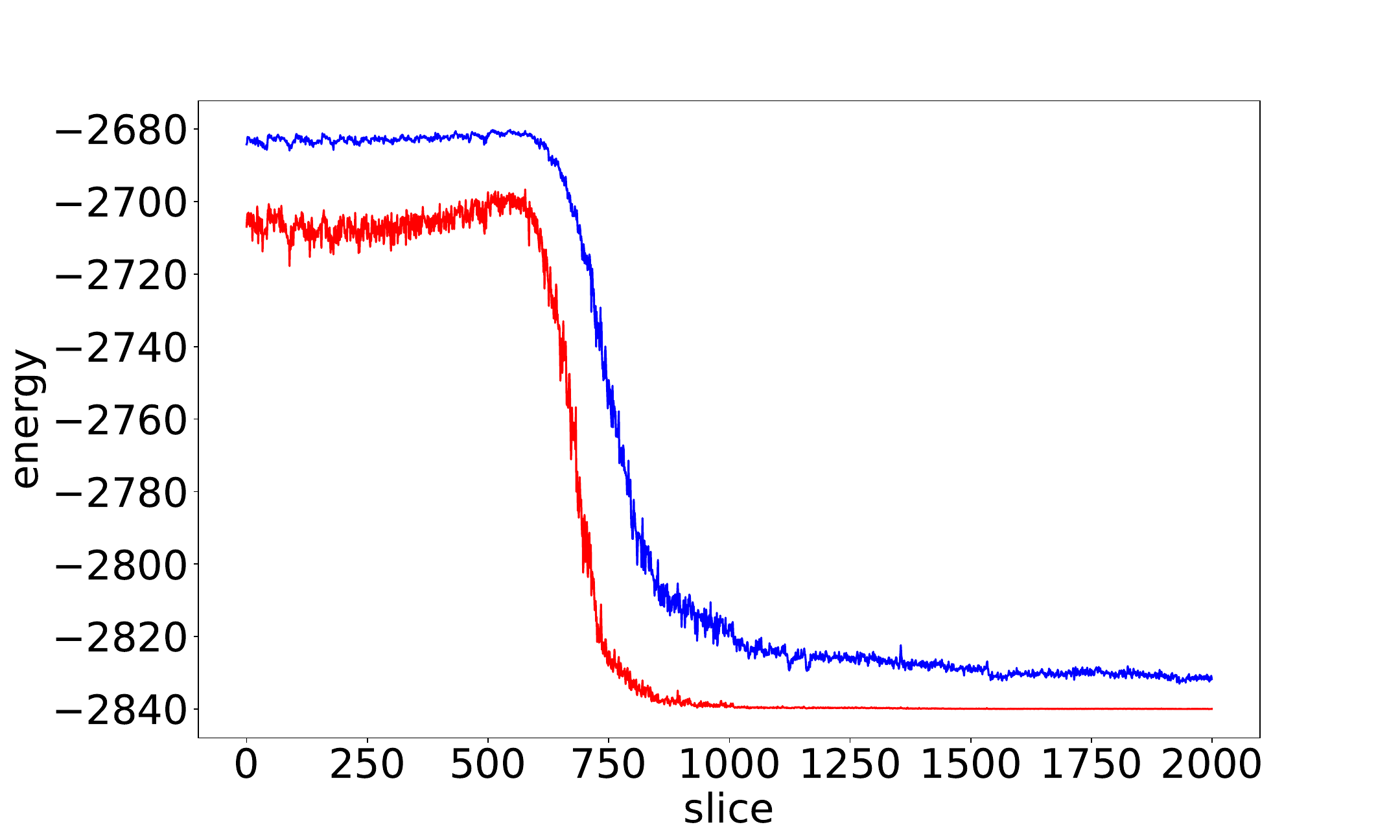}
    \caption{Evolution of minimum energy states on D-Wave 2000Q for 1000 slices (left) and 2000 slices (right). Mean of all samples (blue) and mean of the lowest $1\%$ energies (red).}
    \label{fig:energies}
\end{figure*}

Figure~\ref{fig:energies} displays results for the Chimera Ising model of Algorithm~\ref{algo:genetic} with $1000$ and $2000$ slices. We observe that, initially, the energies roughly stay constant up to around a third of the anneal time. At that point, a continuous reduction in energies sets in, during which the energy of the current state becomes better and better with every slice. Roughly halfway through the anneal the energies stabilize.

We applied the freezeout point estimation algorithm described in Section~\ref{sec:measuring_freezeout}, but it failed to produce a freezeout point estimate for both $1000$ and $2000$ slices, which indicates that the distribution of the samples' energies does not follow a Boltzmann distribution.

\subsection{Evolution of the Hamming distance between adjacent slices}
\label{sec:evolution_hamming}
Similarly to Figure~\ref{fig:energies}, we record the Hamming distance between the binary solution vectors (indicating the final measurement of each qubit) of adjacent slices. This allows us to measure in how many bits the solution vectors of adjacent slices differ.

\begin{figure*}
    \centering
    \includegraphics[width=0.49\textwidth]{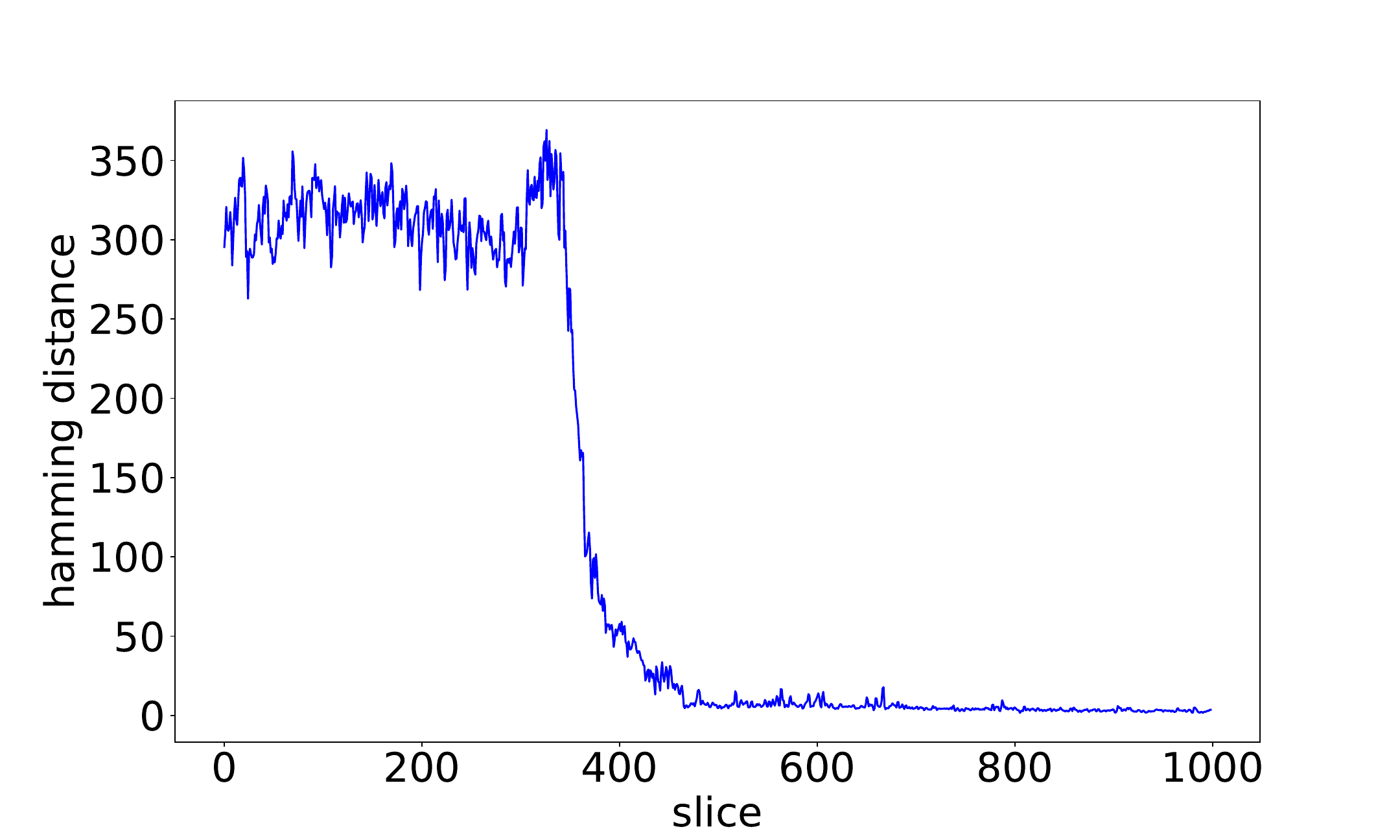}\hfill
    \includegraphics[width=0.49\textwidth]{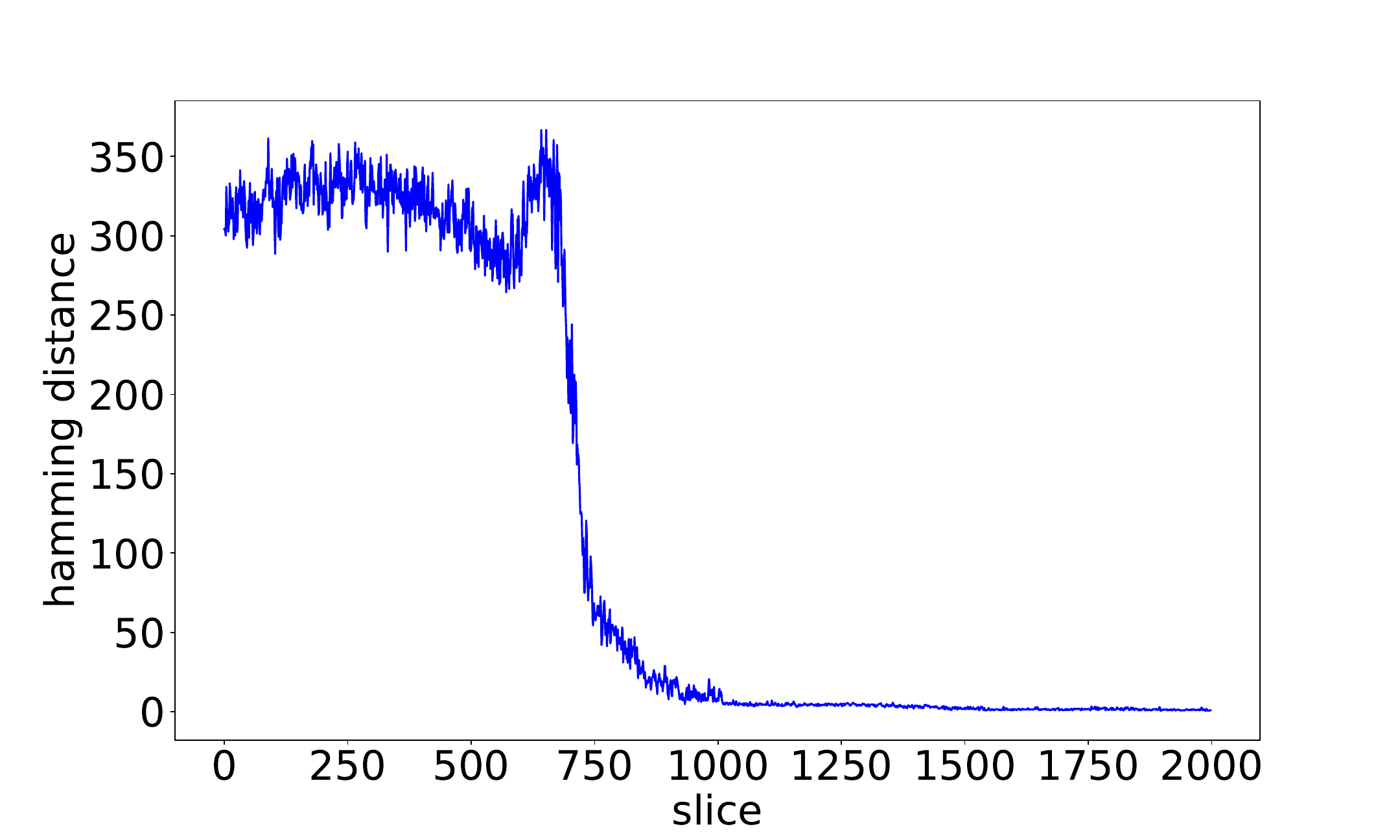}
    \caption{Evolution of the Hamming distance for the quantum states between adjacent slices. Plots show 1000 slices (left) and 2000 slices (right).}
    \label{fig:hamming_distance}
\end{figure*}

Figure~\ref{fig:hamming_distance} shows the evolution of the Hamming distance for the quantum states between adjacent slices for $1000$ and $2000$ microsecond anneals. Interestingly, the shapes of the curves are similar for the two anneal durations. Importantly, they coincide with the shape of the progression of energies in Figure~\ref{fig:energies}. However, the cause of the slight uptick in Hamming distance before the pronounced decrease is unknown.

\subsection{Pausing the anneal}
\label{sec:evolution_pausing}
We are interested in investigating the state evolution when the anneal is paused. In particular, we are interested in observing if the quantum state continues to evolve after the point in Figures~\ref{fig:energies} and \ref{fig:hamming_distance} at which the mean and average best (lowest) $1\%$ energies as well as the Hamming distance stabilize.

For this we select the schedule displayed in Figure~\ref{fig:pausing_schedule}: it has a total anneal time of 2000 microseconds, with a pause inserted at slice 500 out of 1000 slices (microseconds). The pause duration was $1000$ microseconds.

\begin{figure*}
    \centering
    \includegraphics[width=0.49\textwidth]{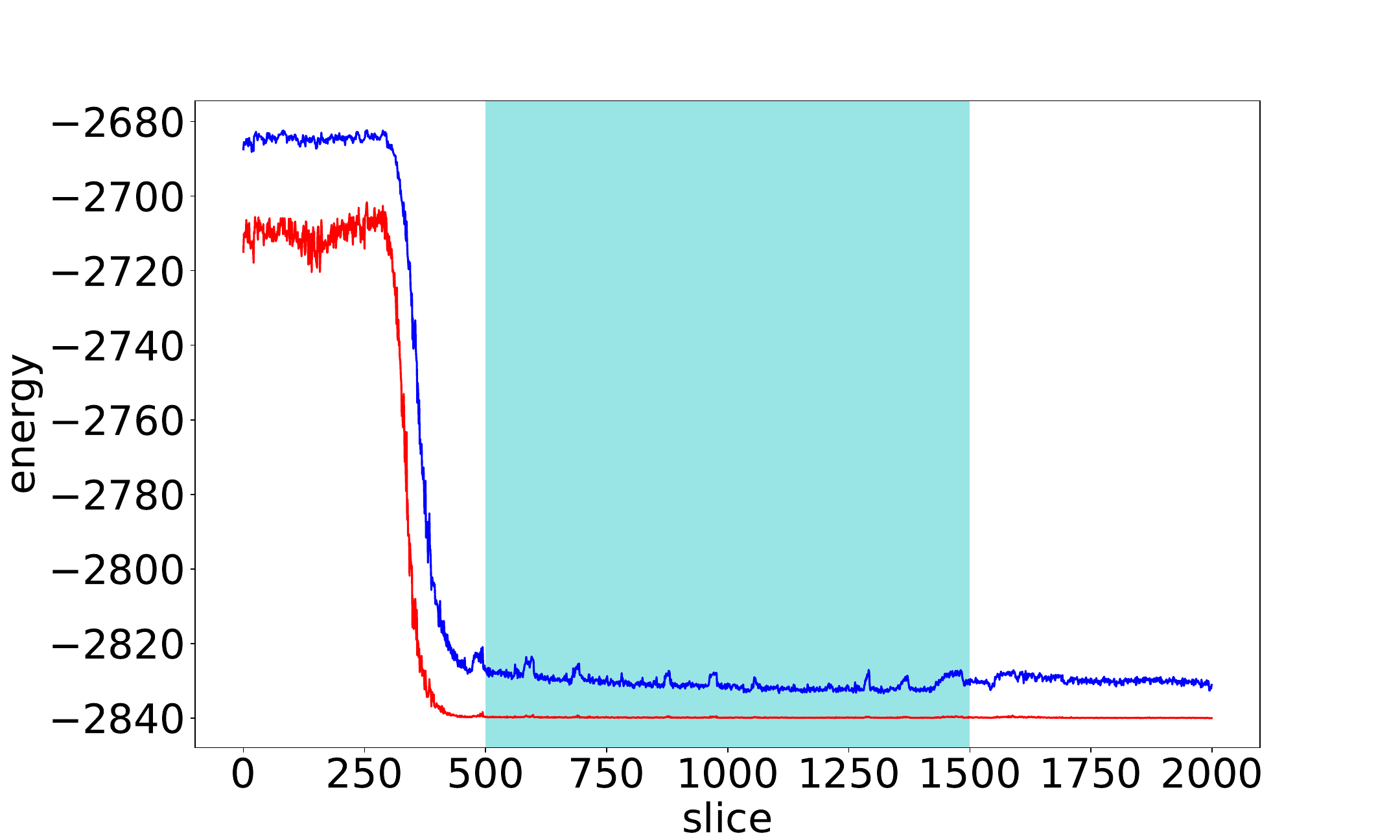}\hfill
    \includegraphics[width=0.49\textwidth]{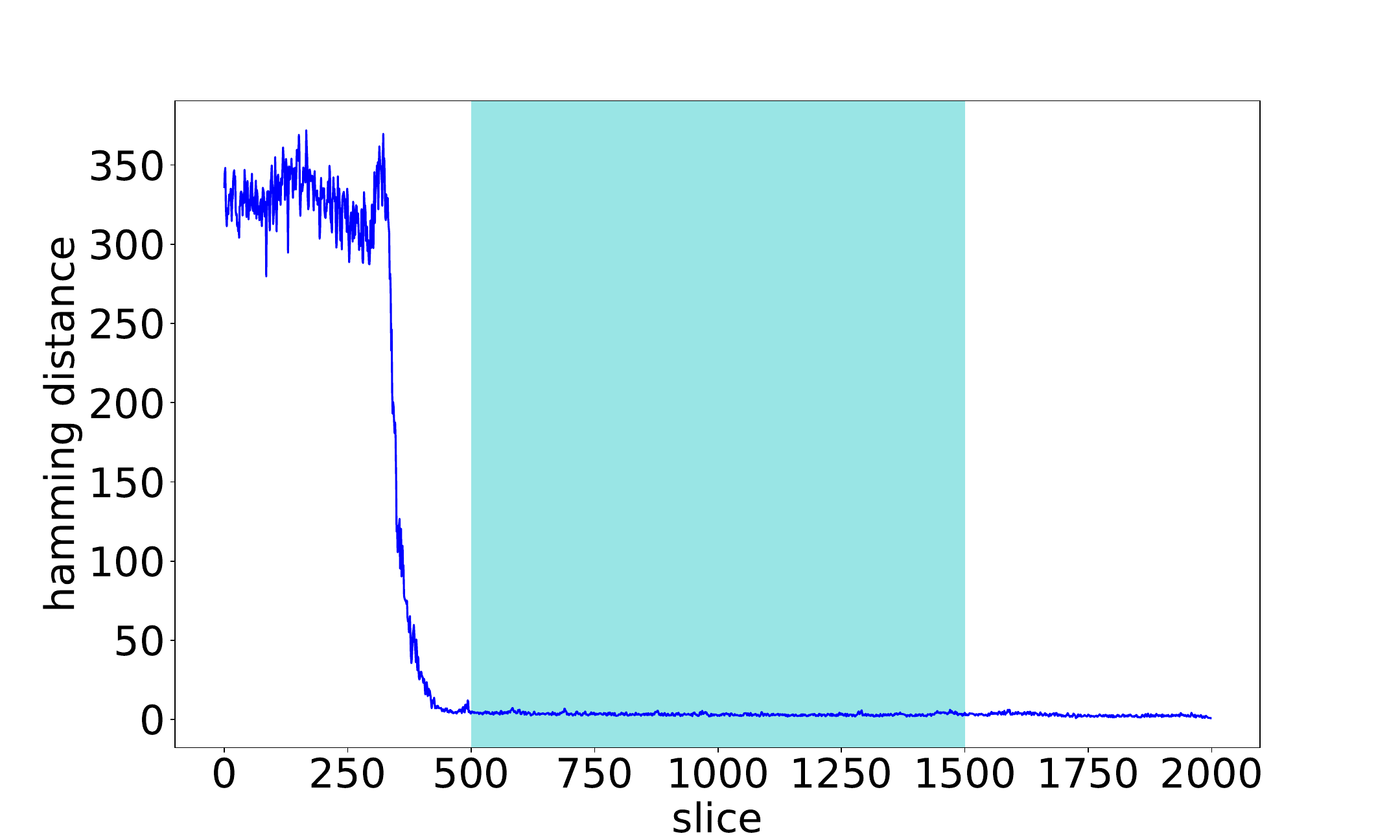}
    \caption{Evolution of energies (left) and Hamming distance (right) during a 2000 microsecond anneal with a 1000 microsecond pause. The pause was inserted at the lower bound of the freezeout point estimate from Figure~\ref{fig:energies}. Pause location is shaded in cyan from 500 to 1500 microseconds.}
    \label{fig:pausing}
\end{figure*}

Figure~\ref{fig:pausing} shows results of this experiment. We observe that, as before, both the energy and Hamming distance of the current states continue to decrease during the pause, which is shaded in light blue. Surprisingly, as soon as the pause is over, we observe a slight uptick in the energy measurements, which then quickly stabilize again towards the end of the 2000 microsecond anneal. The reason of this behavior is unknown, but it could be related to an artifact caused by the D-Wave annealer.

\subsection{Energy evolution for the Maximum Clique problem}
\label{sec:optimizedMaxClique}
We are interested in peering into the anneal process not only for arbitrary (unstructured) Ising problems that conform to the Chimera topology, but also aim to apply the slicing approach to chained problems of practical importance. An important feature of such problems is that they require, before annealing, a minor-embedding onto the Chimera graph, which means they have chains (sets of strongly coupled physical qubits) corresponding to a logical qubit. Chained problems are believed to have a different behavior (i.e., more difficult for quantum annealing) compared to the unchained ones. We select an important NP-hard problem, the Maximum Clique problem \cite{Lucas2014}, to analyze with our slicing method. Since the QUBO for the Maximum Clique problem is of a special form, we employ a modification of Algorithm~\ref{algo:genetic} for the optimization, using $200$ generations. In particular, we start with an initial population of Maximum Clique QUBOs, and every time a crossover or mutation step is performed, we directly work on the edges of the graph for which the Maximum Clique problem is computed. This ensures that, after crossover and mutation, the population of QUBOs continues to contain only those QUBOs that actually correspond to a Maximum Clique problem. In this way, we can find a QUBO of the type of eq.~\eqref{eq:hamiltonian} that corresponds to an actual Maximum Clique problem, while also ensuring a large difference between the energies for 1 microsecond and 1000 microsecond anneals, thereby reducing the distortion caused by quenching.

To sample initial graphs $G=(V,E)$ for which the Maximum Clique problem is solved, we draw random graphs with $|V|=65$ vertices and a density (edge probability) that is uniformly sampled in $[0.2,0.8]$. When embedding the resulting Maximum Clique QUBOs, we employ a fixed chain strength of $2$. When running the genetic algorithm, we employ the default choices of Section~\ref{sec:GA_parameter_choice}, apart from setting $p_\text{mut}=0.01$ due to the fact that we have a smaller number of parameters in this setting.

\begin{figure*}
    \centering
    \includegraphics[width=0.49\textwidth]{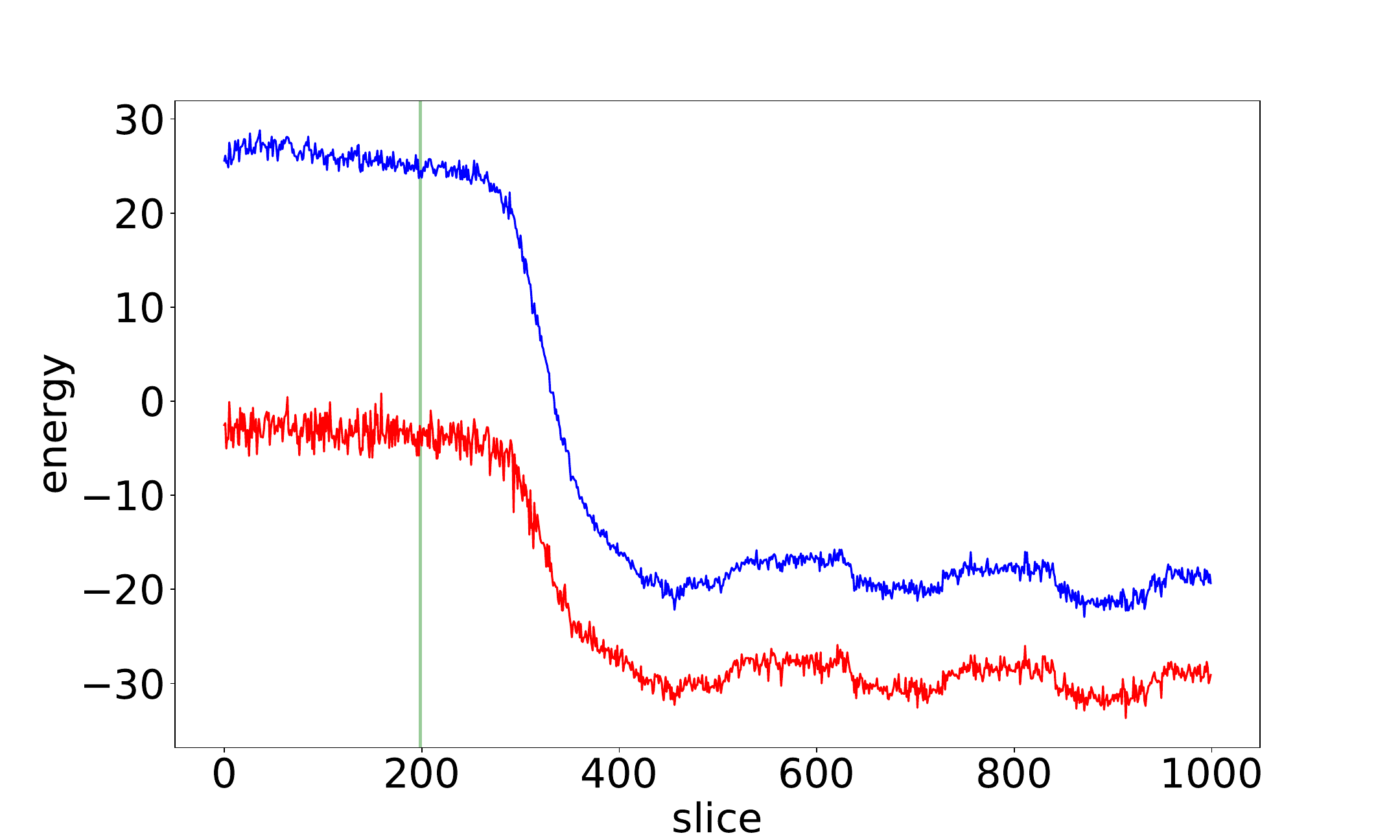}\hfill
    \includegraphics[width=0.49\textwidth]{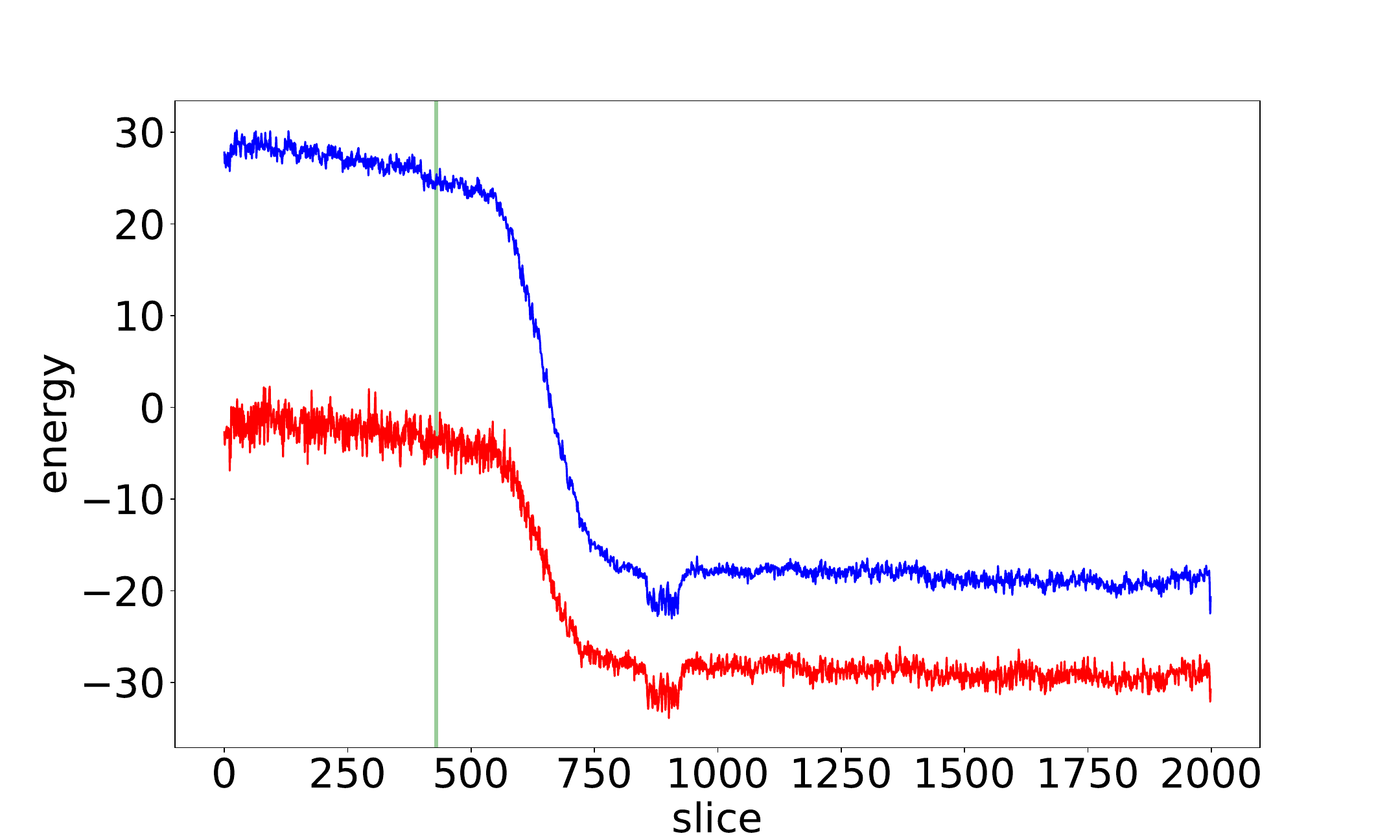}
    \caption{Evolution of energies for the optimized Maximum Clique QUBO using 1000 slices (left) and 2000 slices (right). Mean of all samples (blue) and mean of the lowest $1\%$ energies (red). Estimate of the freezeout point shaded in green.}
    \label{fig:opt_maxclique_energies}
\end{figure*}

Figure~\ref{fig:opt_maxclique_energies} shows an interesting behavior of the Maximum Clique QUBO returned by the modified Algorithm~\ref{algo:genetic}. As seen before, at the beginning of the anneal, both the energies as well as the Hamming distance between adjacent slices slightly decrease before the pronounced decrease sets in. Notably, there seems to be some sort of minimum right after this pronounced decrease for both energy and Hamming distance, and the energy measurements for the QUBO continue to be volatile (oscillatory).

We compute a freezeout point estimate using the algorithm of Section~\ref{sec:measuring_freezeout}. We observe that the result indicates that the freezeout occurs earlier than suggested by our slicing plots. Note that if the quenching used by our slicing algorithm modifies the position in the slicing plot where the energy evolution becomes negligible, it will move the freezeout estimate to an earlier point, but not to a later one. This means that the discrepancy in the position of the freezeout point estimate is due mostly to inaccuracies in the algorithm of Section~\ref{sec:measuring_freezeout}, since the slicing algorithm can only underestimate the freezeout point, but not overestimate it.

\begin{figure}
    \centering
    \includegraphics[width=0.49\textwidth]{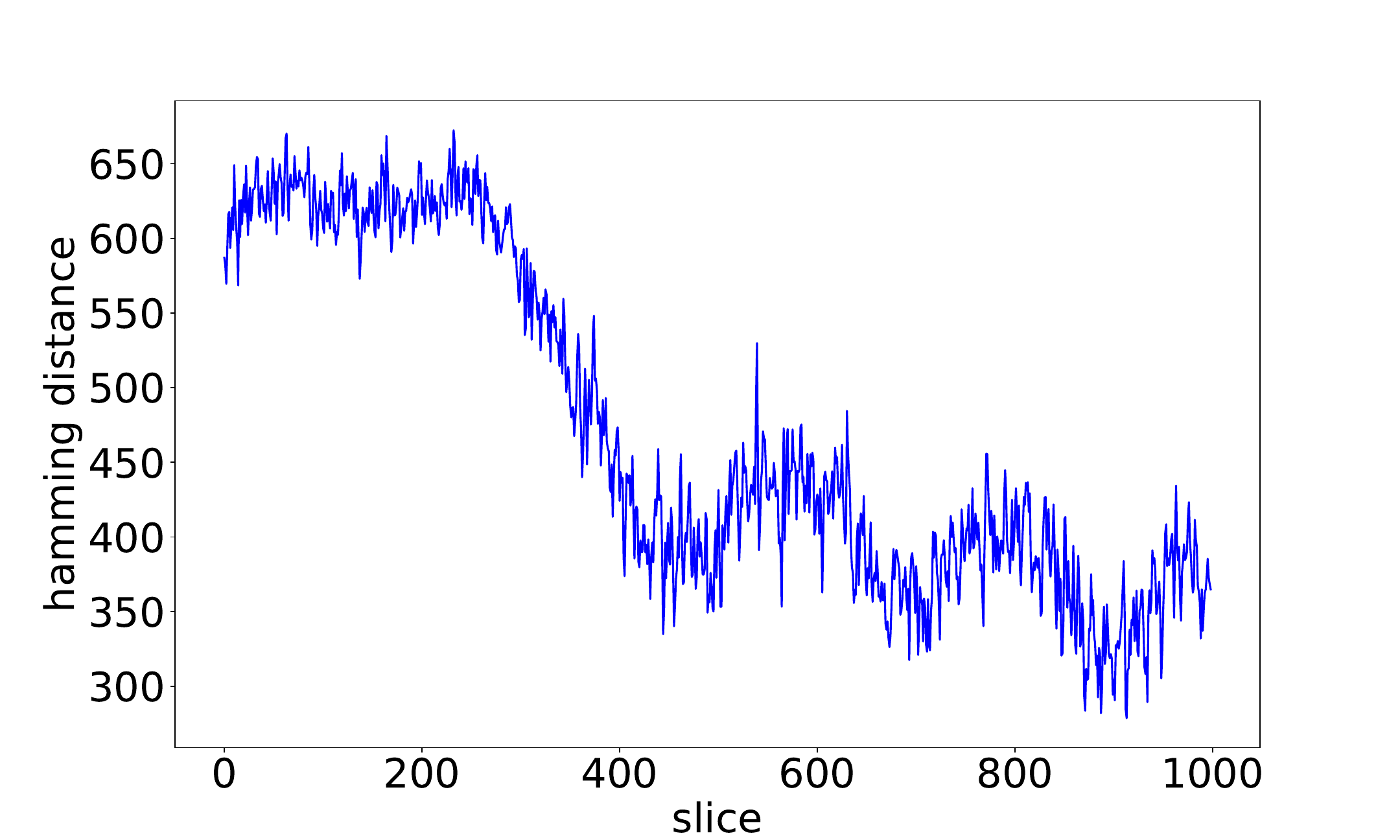}\hfill
    \caption{Evolution of the Hamming distance for the Maximum Clique QUBO using $1000$ slices.}
    \label{fig:opt_maxclique_hamming}
\end{figure}

Similarly to Figure~\ref{fig:opt_maxclique_energies}, Figure~\ref{fig:opt_maxclique_hamming} displays the evolution of the Hamming distance between adjacent slices during a $1000$ microsecond anneal. We observe that, similarly to the energy evolution in Figure~\ref{fig:opt_maxclique_energies}, the Hamming distance first stays relatively constant before a pronounced drop occurs. Two observations are noteworthy. First, we again observe a minimum after the drop. Second, we observe even more pronounced oscillations than in Figure~\ref{fig:opt_maxclique_energies}.

\subsection{Quasi-freezeout point}
\label{sec:spline}
We observed in the previous subsections that each of the shapes of the energy evolution plots follows roughly the same pattern: After an initial phase, in which the energy of the state stays mostly constant (Phase 1), a pronounced decline sets in (often at around one third of the anneal time). The phase of the steep decline (Phase 2) roughly lasts until about halfway during the anneal, after which the energies stabilize quickly. This is followed by Phase 3, which consists of a relatively long stretch of time until the end of the anneal, in which the energy is roughly constant or is decreasing very slowly. Assuming the slope of the line that approximates the energy slicing plot of this phase is not too steep (compared with a user specified threshold), we call the point of the anneal process at which the third phase starts a \textit{quasi-freezeout point (QFP)}.

\begin{figure*}
    \centering
    \includegraphics[width=0.49\textwidth]{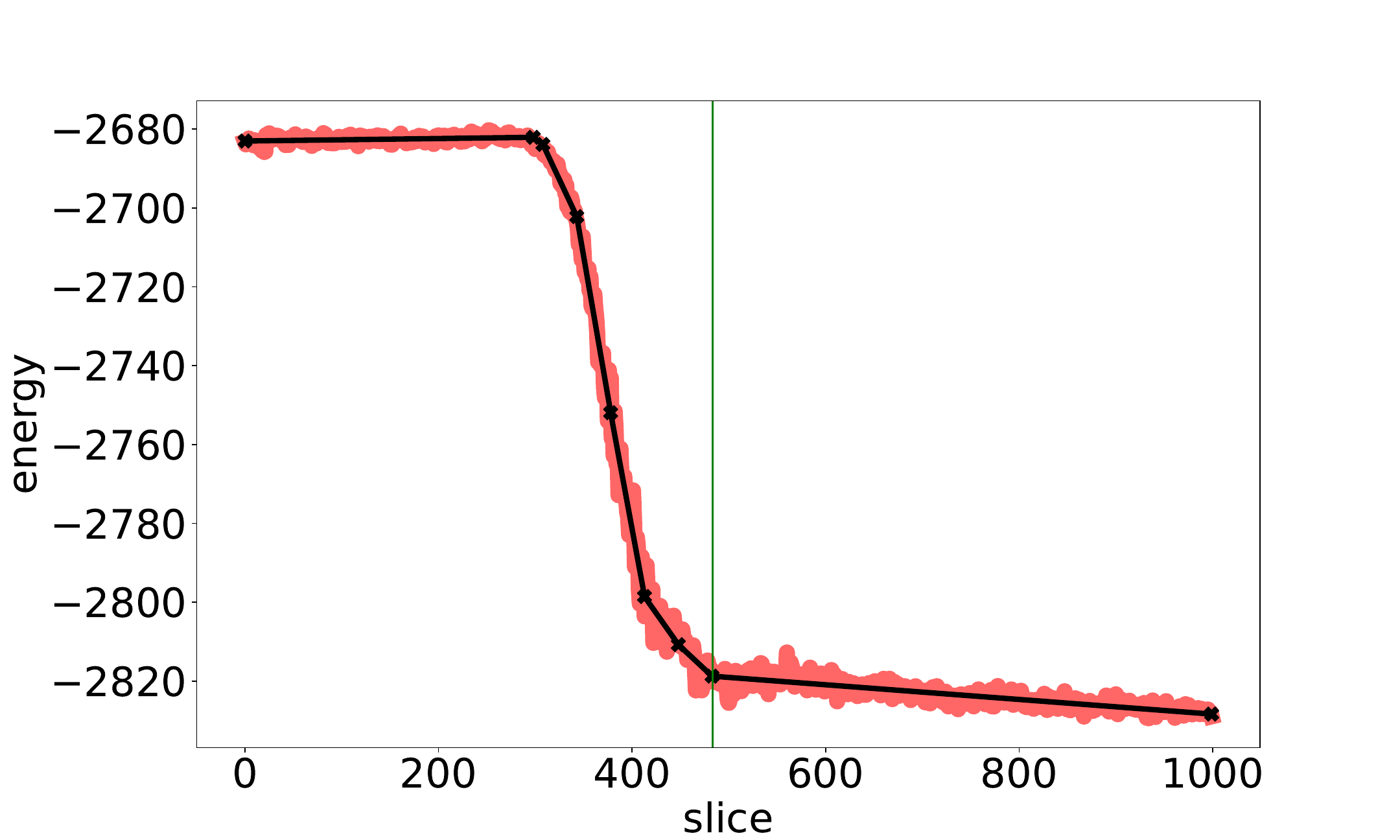}\hfill
    \includegraphics[width=0.49\textwidth]{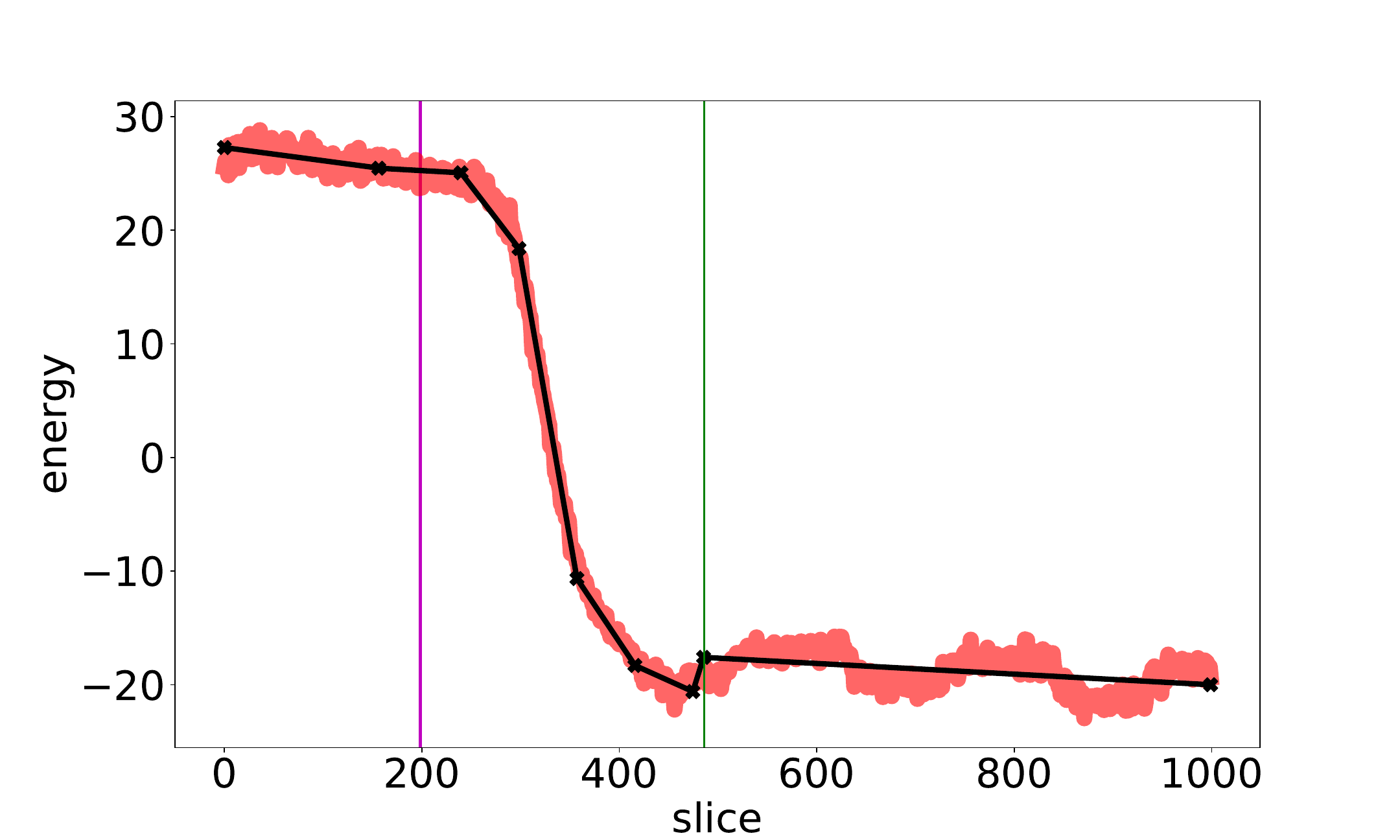}\\
    \includegraphics[width=0.49\textwidth]{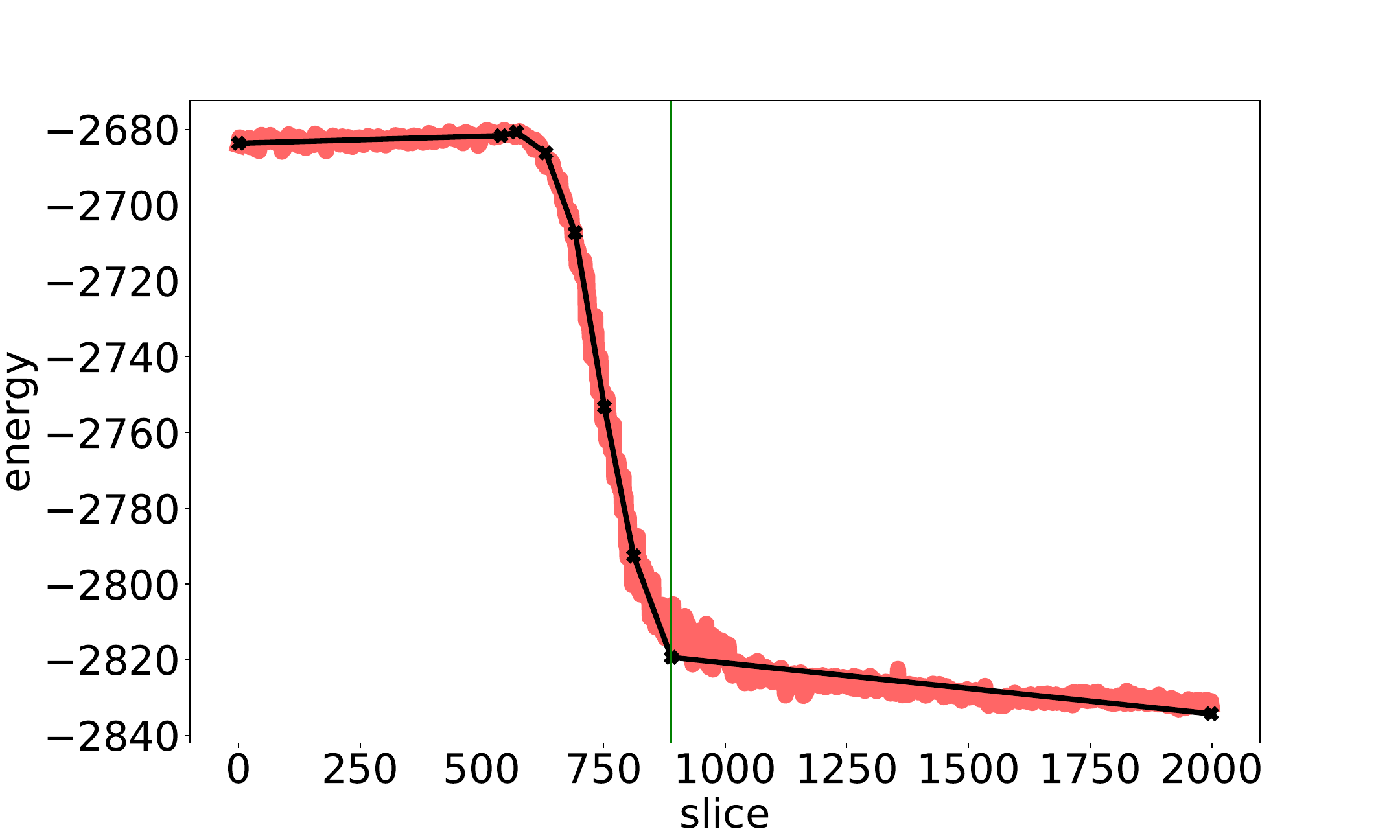}\hfill
    \includegraphics[width=0.49\textwidth]{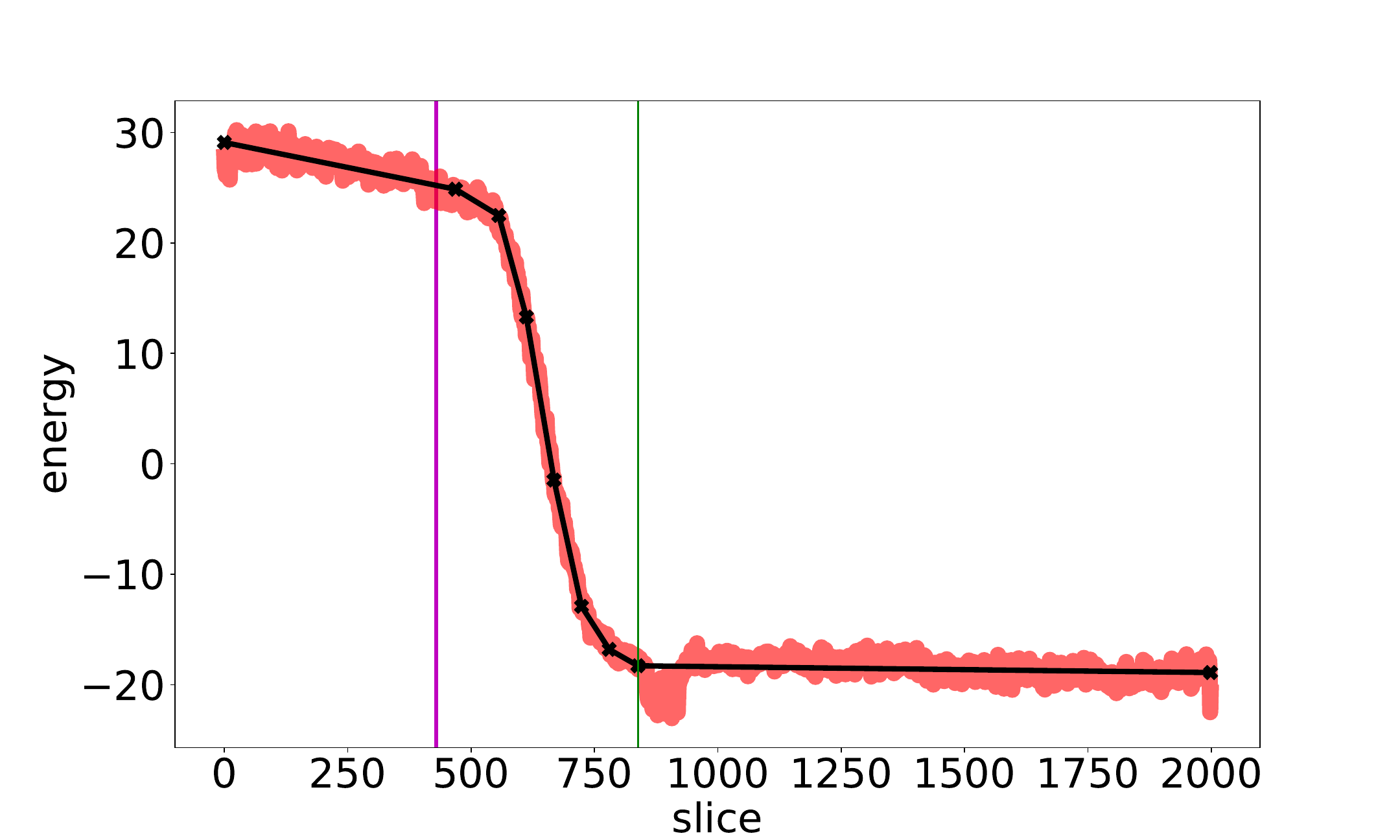}
    \caption{Chimera Ising problem (top left) and Maximum Clique problem (top right), both with anneal duration $1000$ $\mu$s and found using Algorithm~\ref{algo:genetic}. The same problems, but with anneal duration 2000 $\mu$s, are shown in the bottom row. Green vertical lines show the QFP found by the spline method. Purple vertical lines show the freezeout point estimates computed by the method of Section~\ref{sec:measuring_freezeout} (that method could not determine any freezeout point estimates for the Chimera Ising problem).
    \label{fig:spline_freezeout}}
\end{figure*}

In order to divide the energy slicing plot into the three phases described above, we fit a polyline (degree-one spline) using the Bayesian Optimization package of \cite{Bayesian-Optimization}, which consists of three main segments, corresponding to the three phases, and possibly a small number of shorter connecting segments, depending on the graph, to each slicing energy plot. Figure~\ref{fig:spline_freezeout} shows results of our fit. We consider both the Chimera Ising model of Section~\ref{sec:random_vs_optimized} and the Maximum Clique problem of Section~\ref{sec:optimizedMaxClique}, obtained with Algorithm~\ref{algo:genetic}.

Fitting a spline to the slicing energies gives our method two advantages: First, the existence of a QFP can be made dependent on a threshold for the slope set by the user, which typically varies based on the problem being solved, the annealer hardware, and the purpose of the analysis. If the slope of the third segment is zero (i.e., not significantly different from zero), or less than the user-defined threshold, we conclude that the freezeout must have occurred at the intersection of the penultimate and ultimate segment, that is, the point at which the third segment begins. In this case, we can use the QFP as an approximation of the freezeout point. In our case, we consider a system "frozen out" if the slope of the last spline segment is less than $10$ degrees. Second, if the slope is greater than the user-defined threshold, we conclude that the system has not frozen out prior to the end of the anneal. This corresponds to the case where a freezeout point does not exist for the system.

In Figure~\ref{fig:spline_freezeout}, we examine the QFP found for two problems: the Chimera Ising model from \ref{sec:random_vs_optimized} (i.e., not corresponding to any particular NP-hard problem), and the one for the Maximum Clique problem from \ref{sec:optimizedMaxClique}. Both are found with Algorithm~\ref{algo:genetic}. We display the fitted spline segments, and indicate the QFP determined by the last spline with a green line. Additionally, we compute a freezeout point estimate with the method of Section~\ref{sec:measuring_freezeout}. We observe that for the Chimera Ising problem, the technique of \cite{Benedetti2016} does not work, and that it estimates the freezeout point to be before the point at which the energies stabilize for the Maximum Clique problem (purple vertical lines, right column). The QFP is indicated with a vertical green line and could be a sensible indicator for the freezeout of the system.

The 3-tuple of the slopes of the three segments corresponding to Phases 1, 2, and 3, together with the QFP,  could  be possibly used as a signature vector characterizing the anneal evolution for each problem. Since we have looked at only two problems, we cannot suggest any rules that link such a signature with characteristics of an individual problem, but it may be a topic of future reseach.

The precise slope estimates (in degrees) in each of the subfigures of Figure~\ref{fig:spline_freezeout} can be found in Table~\ref{table:slopes}.

\subsection{Proportion of chain breaks}
\label{sec:chainbreaks}
Our slicing technique allows us to also look at the progression of individual qubits during the anneal (see Section~\ref{sec:freezeout_chimera}). In particular, for the Maximum Clique QUBO we can investigate how its chains on the D-Wave Chimera graph evolve.

\begin{figure}
    \centering
    \includegraphics[width=0.49\textwidth]{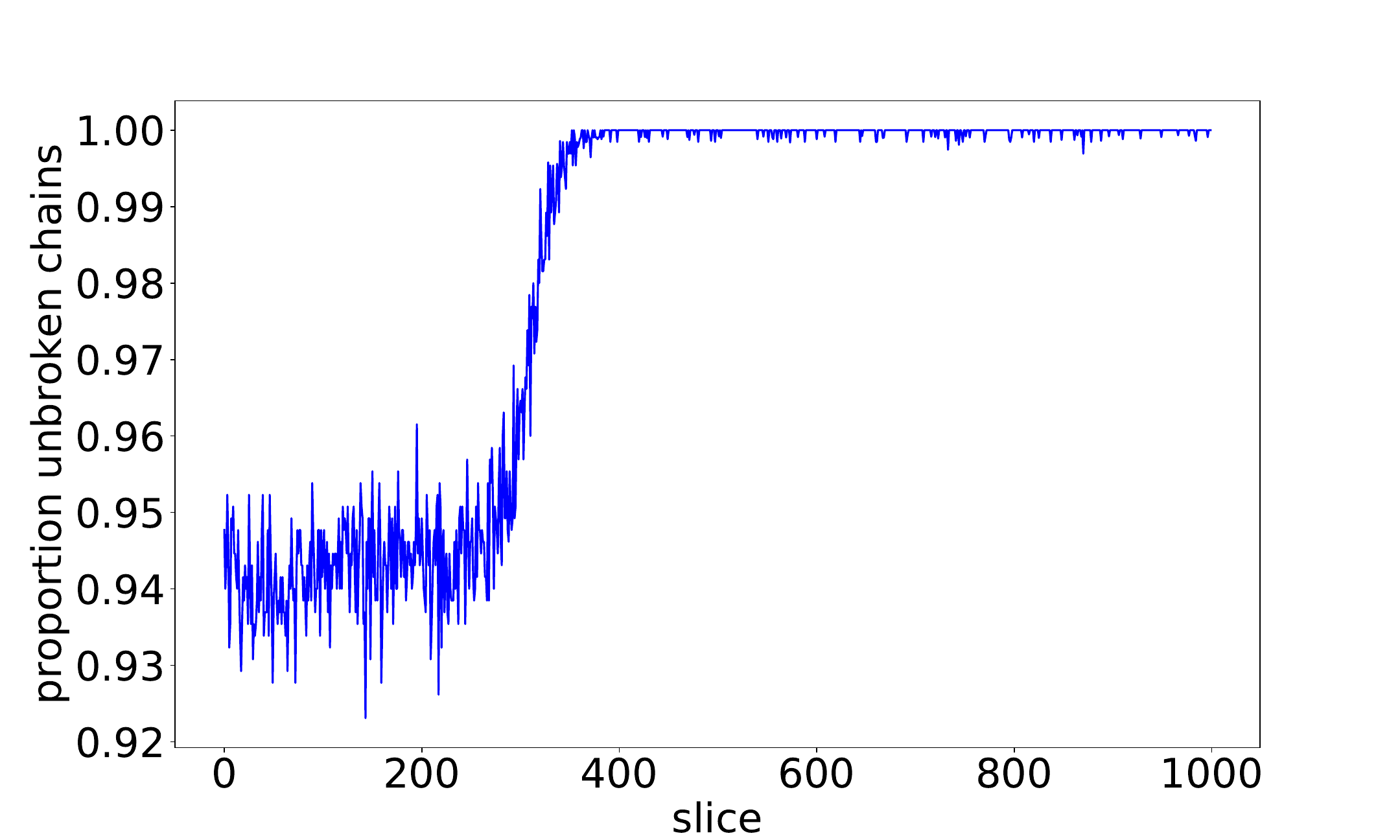}
    \vspace{-2em}
    \caption{Proportion of unbroken chains for the Maximum Clique QUBO using 1000 slices.}
    \label{fig:opt_maxclique_chain_breaks}
\end{figure}

\begin{table*}[t]
\centering
\caption{Spline segment slopes (in degrees) for Figure~\ref{fig:spline_freezeout}\label{table:slopes}\vspace{-1em}}
\begin{tabular}{|l|l||l|l|l|}
    \hline
    Problem ~ & anneal time & Phase 1 ~ & Phase 2 & Phase 3~ \\
    \hline
    Chimera Ising & $1000$ $\mu$s & ~~~3.45 & -84.21 & -7.07\\
    Chimera Ising & $2000$ $\mu$s & ~~~3.0 & -84.54 & -9.71\\
    Maximum Clique & $1000$ $\mu$s & -16.92 & -84.42 & -2.32\\
    Maximum Clique & $2000$ $\mu$s & -17.46 & -84.3 & -1.31\\
\hline
\end{tabular}
\end{table*}

Figure~\ref{fig:opt_maxclique_chain_breaks} displays the proportion of unbroken chains (where an unbroken chain is defined as a chain whose qubits all take the same value) as a function of the anneal slice for a $1000$ microsecond anneal. We observe that initially, up to around one third of the anneal, roughly $95\%$ of all chains are unbroken. Towards one third of the anneal time, roughly coinciding with the drop in energies, all of the chains become unbroken.

\subsection{Determining quasi-freezeout points at individual-qubits level}
\label{sec:freezeout_chimera}
We can use our methodology from Section~\ref{sec:slicing} to track when individual qubits of D-Wave ``freeze" during the anneal. To this end, we again employ the Ising or QUBO model obtained with the genetic algorithm of Section~\ref{sec:genetic_algo}, and read out the value of each qubit at each of the $1000$ slices during a $1000$ microsecond anneal.

In this way, we can track from what slice onwards the value of each of the qubits remained in the state it was upon readout at the full anneal duration. We define this timepoint as the \textit{quasi-freezeout point (QFP)} for that individual qubit. The bitstrings we query for each slice are the ones corresponding to the best energy.

\begin{figure*}
    \centering
    \includegraphics[width=0.49\textwidth]{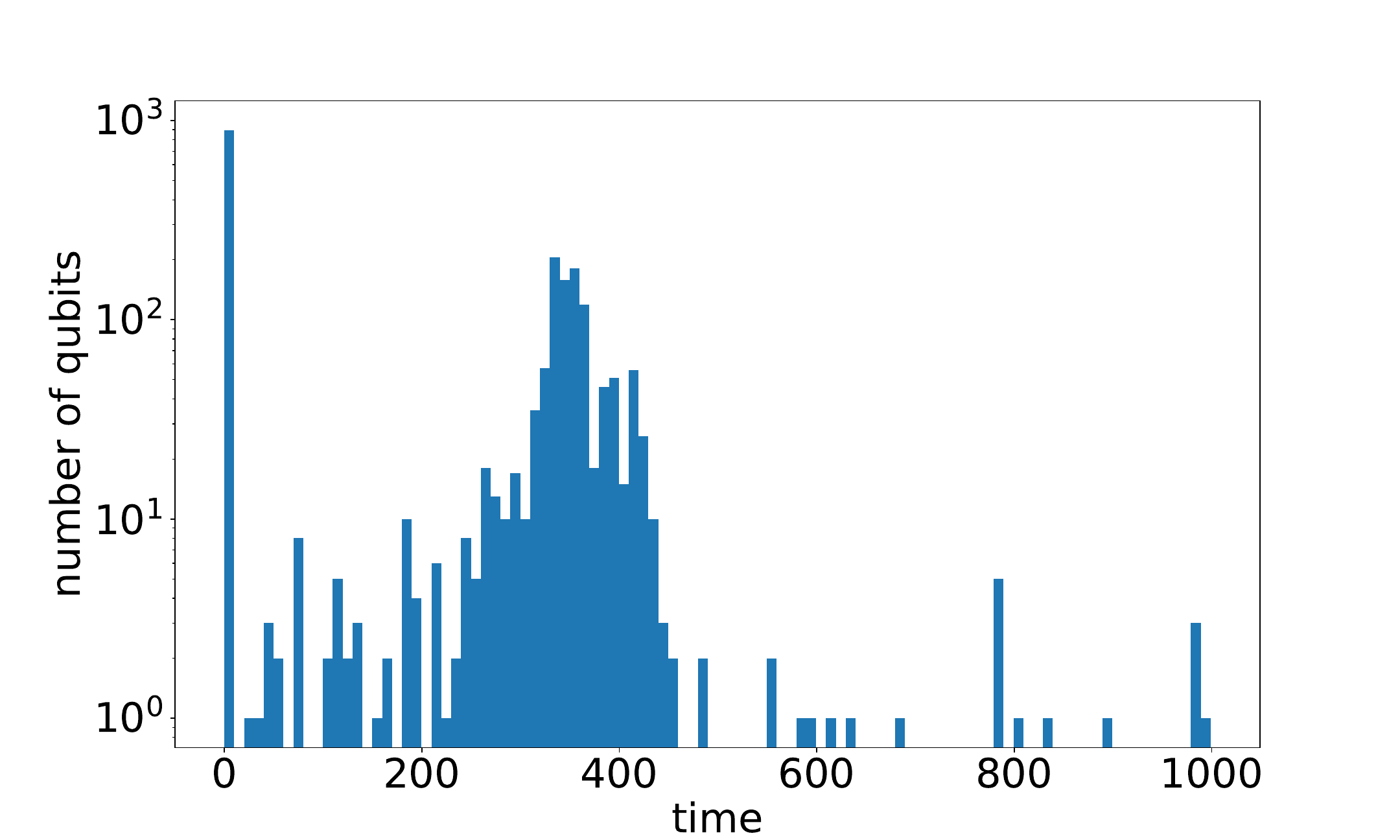}
    \includegraphics[width=0.49\textwidth]{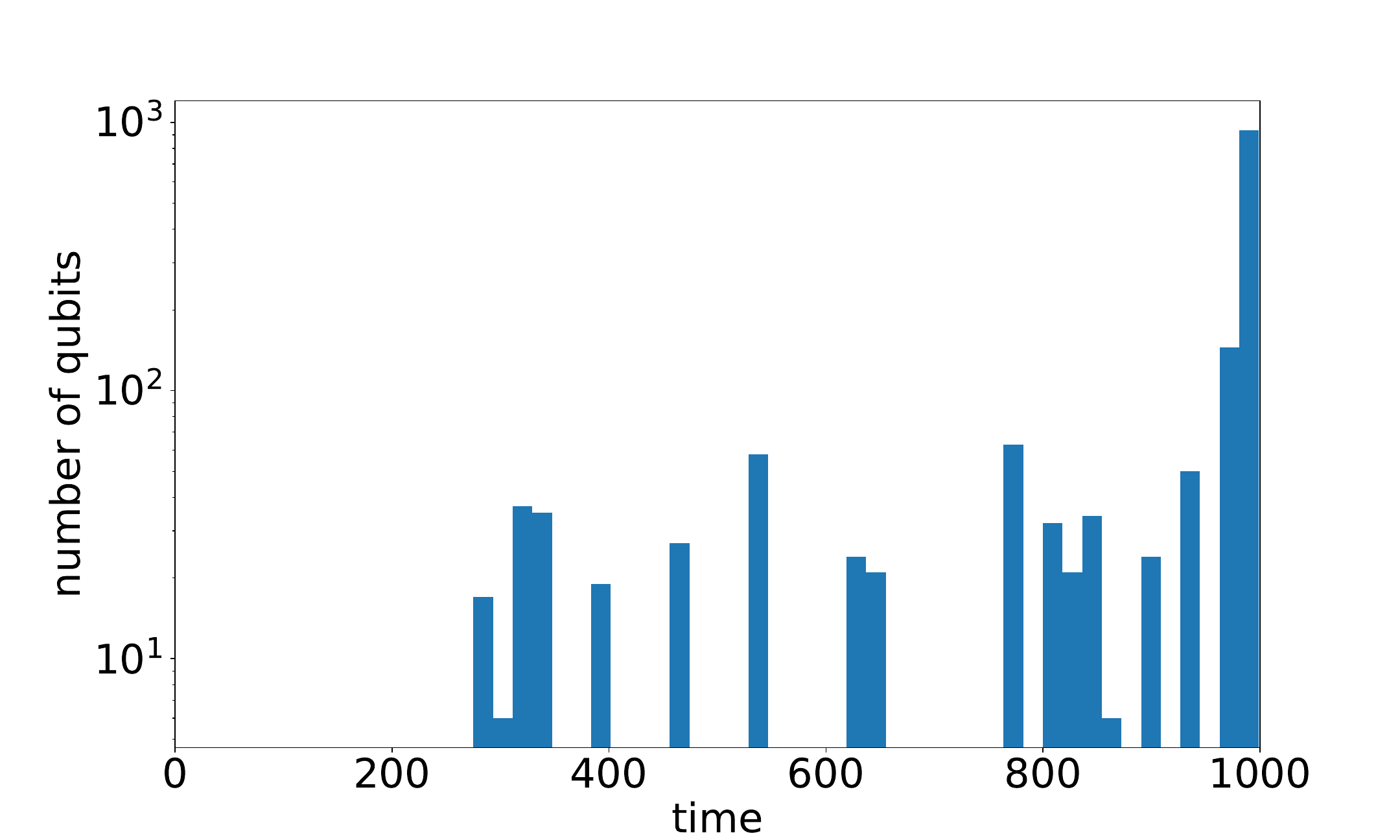}
    \vspace{-0.4em}
    \caption{Histograms showing the number of frozen out qubits at each slice. Left histogram shows the Chimera Ising model, and right histogram shows the Maximum Clique problem, obtained with Algorithm~\ref{algo:genetic}.}
    \label{fig:freezeout_chimera}
\end{figure*}
Figure~\ref{fig:freezeout_chimera} shows results of this experiment for the Chimera Ising (having $2032$ variables) and the Maximum Clique problems (having $1555$ variables) obtained with Algorithm~\ref{algo:genetic}. Each of the two histograms shows the number of qubits that freeze out at a particular point in time (or slice number) during the anneal.

We observe that, for the Chimera Ising model, the histogram correlates with the slicing diagrams we have seen earlier. In particular, at the start of the anneal, not many qubits freeze out. Roughly at one third of the anneal duration, when we observe the pronounced decrease in the slicing energies, a majority of qubits freeze out. The one exception, which, on the surface, does not make sense, is the large number ($\approx\! 900$) of qubits that have frozen out at slice 1. Actually, this gives out some useful information, since these qubits must have frozen because of the quench--they have attained their optimal values as a result of the quench done as part of the slicing at 1 $\mu$s time. It cannot be explained by the 1 $\mu$s anneal that precedes that quench because, at $s=0.001$, the value of the function $B(s)$ from Figure~\ref{fig:anneal_schedule} is practically zero, so the annealer does not have enough information about the QUBO or Ising model of eq.~\eqref{eq:hamiltonian} in order to compute these optimal values. Hence, we can conclude that the distortion caused by the the quenches is expressed by having the values of $\approx\! 900$ qubits fixed, while the remaining $\approx\! 1100$ qubits determine the reduced QUBO or Ising model whose anneal progression we observe on the slicing diagrams for the Chimera Ising problem.

For the Maximum Clique problem, the histogram on the right of Figure~\ref{fig:freezeout_chimera} tells a different story. No qubit freezes out before slice 275, and after that, qubits freeze out at seemingly random times until almost the end of the anneal. This is consistent with the slicing diagram, which shows that the quantum state energy keeps on decreasing slowly until the end of the anneal. The increased number of frozen qubits at the very end is due to the fact that the annealing is soon ending and there is not enough time (slices) to allow for the qubits to flip again. The fact that the freezeout estimates of the qubits are spaced out, with longer intervals between them, and that several qubits are estimated to freeze out at the same time is due to the chained nature of the Maximum Clique problem. Qubits representing a logical qubit are chained together and thus, typically, either a logical qubit flips, or a part of a broken chain flips. Finally, the lack of frozen qubits at slice 1 indicates that, for the Maximum Clique problem, the quench did not result in such a big distortion as happened for the Chimera Ising problem, and the slicing result is more reliable. On the other hand, the slicing diagrams for both the Chimera Ising and optimized problems have similar shapes, which may indicate that, despite the distortion caused by the quench, the Chimera Ising plots still yield useful information.

\section{Discussion}
\label{sec:discussion}
This article is a first attempt to explore how the state of a quantum annealer evolves during annealing, and a first step in the development of methods that allow us to get further insights into the (unobservable) anneal process. To the best of our knowledge, such work has not been presented previously in the literature.

We develop a novel method we refer to as slicing, based on the quenching control feature of D-Wave 2000Q, to estimate the states of individual qubits at any point during the anneal. Using this technique, we dissect the anneal process and monitor how the energy of the state evolves. The slicing plots allow us to track when energies or qubit bit flips stabilize during the anneal, which can be used as an alternative to estimate the freezeout point, similarly to the method of \cite{Benedetti2016}. We summarize our findings as follows:
\begin{enumerate}[wide]
    \item We show that an optimized Ising or QUBO model, computed with a genetic algorithm, exhibits a much more pronounced evolution during the anneal than a random problem instance.
    \item Using optimized Ising or QUBO models, we observe that the evolution of energies and Hamming distances between adjacent states follows a similar pattern in the experiments we conducted. Initially, the energy does not decrease considerably. At roughly a quarter of the anneal, a pronounced decrease sets in, which correlates with a reduction in the number of bit flips. At around the midpoint of the anneal, the energy stabilizes at around the energy value that D-Wave returns as a solution at the end of the anneal. During that phase, the number of bit flips likewise stabilizes at a constant level (i.e., the state still changes from slice to slice), which could reflect the noise in the machine.
    \item Our technique provides an alternative to approximate the freezeout point, similarly to \cite{Benedetti2016}. We introduce the notion of a quasi-freezeout point based on the data collected during slicing, which can be used to approximate the freezeout point in cases it cannot be computed by other methods, and suggest an algorithm for its computation.
    \item We demonstrate that the quantum state still keeps on evolving even during a pause in the anneal process.
    \item For the case of Maximum Clique problem, we show that the proportion of broken chains is low at the start of the anneal and quickly decreases to zero, meaning that all chains are unbroken early on in the anneal.
    \item We show that the Chimera Ising and the Maximum Clique problem have quite different behaviors. While, in the former, a large number of qubits freeze very early during the anneal, for the latter no qubits freeze out during the first quarter of the anneal, after which qubits obtain their final values at roughly the same rate until  the end of the anneal.
\end{enumerate}
By applying the slicing method to other random Chimera Ising problems, as well as other realizations of the optimized QUBO stemming from the genetic algorithm, we confirm that the results of this article hold true in greater generality. A multitude of further research avenues are possible:
\begin{enumerate}[wide]
    \item Improving the genetic algorithm implementation by including possibly advantageous characteristics of genetic algorithms such as multiple individual crossovers, adaptive mutation and crossover rates, and different types of elitism.
    \item Slicing problems using other types of anneal schedules such as reverse annealing schedules.
    \item Slicing other problems, including chained optimization problems (e.g., Maximum Cut and Graph Partitioning), and those that are especially difficult for D-Wave to solve.
    \item Investigate for random Ising problems if the set of qubits that freeze out early/late changes. If we consistently observe the same qubits freezing out early/late, we might deduce biases or other properties of the D-Wave machine.
    \item Finally, if future hardware advances allow quenching to become much shorter, e.g., in the low nano-second range, our methods will be able to produce more accurate results.
\end{enumerate}

\section*{Acknowledgment}
This work was supported by the US Department of Energy through the Los Alamos National Laboratory and by the Laboratory Directed Research and Development program of Los Alamos National Laboratory under project numbers 20190065DR and 20180267ER. Los Alamos National Laboratory is operated by Triad National Security, LLC, for the National Nuclear Security Administration of U.S. Department of Energy (Contract No.~89233218CNA000001).


\end{document}